\documentclass[aip,pop,numerical,reprint]{revtex4-1}

\usepackage{dcolumn}
\usepackage{bm}
\usepackage{graphicx}
\usepackage{indentfirst}
\usepackage{units}
\usepackage{color}
\usepackage[hidelinks,colorlinks,breaklinks=true]{hyperref}
\hypersetup{
  colorlinks=true,    
  linkcolor=blue,     
  citecolor=blue,     
  filecolor=magenta,  
  urlcolor=blue,
  bookmarksdepth=4
}
\usepackage[nameinlink,capitalise]{cleveref}
\usepackage[english]{babel}

\begin{document}

\title{
Weakly turbulent plasma processes
in the presence of inverse power-law
velocity tail population
}

\author{S. F. Tigik}
\affiliation{Instituto de F\'{\i}sica, Universidade Federal
  do Rio Grande do Sul, 91501-970 Porto Alegre, RS, Brazil}
\email{sabrina.tigik@ufrgs.br}
\author{L. T. Petruzzellis}
\affiliation{Instituto de F\'{\i}sica, Universidade Federal
  do Rio Grande do Sul, 91501-970 Porto Alegre, RS, Brazil}
\email{larissa.petruzzellis@ufrgs.br}
\author{L. F. Ziebell}
\affiliation{Instituto de F\'{\i}sica, Universidade Federal
  do Rio Grande do Sul, 91501-970 Porto Alegre, RS, Brazil}
\email{luiz.ziebell@ufrgs.br}
\author{P.~H. Yoon}
\affiliation{Korea Astronomy and Space Science Institute,
Daejeon, Korea}
\affiliation{Institute for Physical Science \& Technology,
University of Maryland, College Park, MD 20742, USA}
\affiliation{School of Space Research, Kyung Hee University,
  Yongin, Gyeonggi 446-701, South Korea}
\email{yoonp@umd.edu}
\author{R. Gaelzer}
\affiliation{Instituto de F\'{\i}sica, Universidade Federal
  do Rio Grande do Sul, 91501-970 Porto Alegre, RS, Brazil}
\email{rudi.gaelzer@ufrgs.br}


\begin{abstract}
  Observations show that plasma particles in the solar
  wind frequently display power-law velocity distributions, which can
  be isotropic or anisotropic.  Particularly, the velocity
  distribution functions of solar wind electrons are frequently
  modeled as combination of a background Maxwellian distribution and a
  non-thermal distribution which is known as the ``halo''
  distribution.  For fast solar wind conditions, highly anisotropic
  field-aligned electrons, denominated as the ``strahl'' distribution,
  are also present. Motivated by these observations, the present paper
  considers a tenuous plasma with Maxwellian ions, and electrons
  described by a summation of an isotropic Maxwellian distribution and
  an isotropic Kappa distribution.  The formalism of weak turbulence
  theory is utilized in order to discuss the spectra of electrostatic
  waves that must be present in such a plasma, satisfying conditions
  of quasi-equilibrium between processes of spontaneous fluctuations
  and of induced emission.  The kappa index and relative density of
  the Kappa electron distribution are varied. By taking into account
  effects due to electromagnetic waves into the weak turbulence
  formalism, we investigate the electromagnetic spectra that satisfy
  conditions of ``turbulent equilibrium'', and also the time evolution
  of the wave spectra and of the electron distribution, which occurs
  in the case of the presence of an electron beam in the electron
  distribution.
\end{abstract}

\maketitle

\section{Introduction}
\label{sec:introduction}

Observations made  in the  space environment consistently  show plasma
particles with velocity distributions that have non-thermal tails, and
frequently  with  anisotropies  \cite{Feldman+75,  Pilipp+87,  Lin+95,
Stone+08,  KruckerBattaglia14, Oka+15}.   Characteristically, observed
solar wind  electrons are modeled  by a combination of  the Maxwellian
core population (with energy in the range of tens of eV) and a tenuous
but  energetic  {\it  halo}  distribution that  contains  a  power-law
velocity  distribution in  the suprathermal  range ($\sim10^2$--$10^3$
eV). For  energy range even higher  than that of the  halo population,
that is, for $\sim20$--$200$ keV  range, {\it superhalo} electrons are
also observed \cite{Wang+12}. The halo and superhalo distributions are
often  modeled by  the  {\it  Kappa distribution}  \cite{Vasyliunas68,
SummersT91,   MaceHellberg95,   LeubnerS00,   LeubnerS01,   Leubner02,
Leubner04b,  Wang+12,  Kim+15}.  For  the  fast  solar wind  condition
\cite{McComas+03},  a  field-aligned  electron beam  called  the  {\it
strahl}  is often  observed  to stream  away from  the  Sun. The  {\it
strahl} is  characterized by the similar  energy range as that  of the
halo  electrons. Observations  show that  the number  density of  {\it
strahl}  decreases as  one  moves away  from the  Sun  while the  halo
density  increases  \cite{Maksimovic+05},   but  their combined density
remains constant, being $\sim$ 4$-$5\% of the total density. The energetic
superhalo  electrons  contribute  very  little  to  the  net  electron
content, as their number density amounts to not more than $10^{-6}$ of
the  total electron  density, but  owing to  their high  energy, their
presence     is    evident     in    the     velocity    or     energy
spectrum.          The observations thus suggest that the {\it strahl}
electrons  are  but  a  field  aligned portion of the halo population,
which  are  gradually  pitch-angle  scattered/diffused  back  to  the
isotropic halo by some unknown processes, of which  the  whistler  wave
fluctuations are the prime candidate \cite{VocksMann03, Kim+15}.

The Kappa distribution was introduced to  phenomenologically describe
the  non-thermal   feature  of  the  electron   velocity  distribution
\cite{Vasyliunas68},  but  appears nowadays  in  the  literature in  a
framework    of     non-extensive    thermo-statistical    equilibrium
\cite{LivadiotisBook}. A family of Kappa distributions are used in the
literature,   which   includes    isotropic   or   anisotropic   Kappa
models. Isotropic Kappa distributions are  usually written in terms of
two different  forms, one  which can  be found  in \cite{Vasyliunas68,
SummersT91,  MaceHellberg95}, and  the  other which  can  be found  in
\cite{Leubner02,  Leubner04b}.  These  two different  forms  of  Kappa
distributions have been used by the plasma physics community, and have
been  the subject  of a  number of  theoretical discussions  in recent
years      \cite{HellbergMBKS09,     HapgoodPDD11,      LivadiotisM13,
Livadiotis2015,  LazarFY16}.    In  the   present  paper,  we   use  a
generic form of Kappa distribution,  which in particular cases can
reproduce the  two widely used forms  mentioned above, and use  such a
distribution to describe  the {\it halo} distribution  in the electron
population.

Recently, however, one of us put forth a rigorous theory of Kappa
distribution from the viewpoint of weak turbulence theory, rather
than treating the Kappa distribution as simply a phenomenological
tool \cite{Yoon2012, Yoon2014}. In such a theory it was shown that
a quasi-stationary state of electrons and a spectrum of
electrostatic Langmuir fluctuations form a self-consistent
pair of solutions of the stationary weak turbulence kinetic
equations. A rather remarkable finding is that such a solution
permits only the Kappa distribution as the legitimate solution
but nothing else if the nonlinear interaction terms in the wave
kinetic equation is considered. This finding may explain the
physical origin of the pervasive Kappa-like electron
distribution functions observed in the space environment.
The accompanying Langmuir fluctuation spectrum, according
to the above-referenced papers \cite{Yoon2012, Yoon2014},
is significantly modified from the thermal equilibrium
form of the spectrum in that the long wavelength regime
of the fluctuation spectrum exhibits an inverse power-law
behavior, $\propto k^{-2}$, while for high $k$, the spectrum
approaches a constant value. These findings and discussions were,
however, carried out on the basis of the simplifying assumption
of a single electron species. This was done for the sake of
simplicity. For the actual situation, as overviewed
above, the solar wind electrons are composed of several
components, typically a quasi isotropic Maxwellian core
plus a quasi isotropic halo population, which is often
modeled by a Kappa distribution. In view of this, it is
timely and appropriate to revisit the problem of solar wind
like electron distribution and the associated Langmuir
fluctuation spectrum for multi, or at least, a two-component
electron plasma.

Ideally, one must obtain the electron distribution {\it and}
the Langmuir fluctuation spectrum in a self-consistent manner
without making any assumption at the outset. This is possible
if one makes a simplifying assumption of a single component
electrons \cite{Yoon2012, Yoon2014}. However, if one is to
consider a multiple (or two component) electrons, then the
situation becomes rather complex. Even if one ignores the
nonlinear coupling term, the self-consistent solution for
{\it both} electron distribution {\it and} the Langmuir
spectrum must be obtained by numerical iteration scheme
\cite{Kim+16}. In the present analysis, we are interested in
revisiting the approach taken in \cite{Kim+16} but within
the context of analytical method. In order to reduce the
complexity of the problem somewhat, we approach the problem
by allowing a model two component electron distribution
function and seeking the Langmuir spectrum intensity,
which is consistent with the model electron distribution function.

Thus, in the first part of the present analysis, we will investigate
the spectral form of the electrostatic fluctuation intensity that
exists in a plasma, satisfying equilibrium conditions between
processes related to spontaneous fluctuations and processes
induced by the waves themselves. The analysis is made in the
framework of weak turbulence theory including spontaneous
effects. We consider an unmagnetized plasma with plasma
particles described by velocity distributions,
which are a summation of an isotropic Maxwellian background and
a ``halo'' characterized by isotropic Kappa distributions of
generic form. The analysis to be made under the framework
of weak turbulence theory shows that electrostatic waves, i.e.,
Langmuir ($L$) and ion-sound ($S$) waves, can be naturally
occurring in a plasma as a result of spontaneous and induced
effects. Electromagnetic waves, i.e. transverse waves ($T$),
cannot be generated by these mechanisms, but can appear due
to nonlinear interactions involving other types of plasma waves.

In the second part of the present paper, we also investigate
the generation of electromagnetic waves, and the
possibility of an approximated asymptotic solution for the
spectrum of transverse waves, obtained as the outcome of
nonlinear processes described by weak turbulence theory.
Investigations on the equilibrium spectra of electrostatic
waves and on the spectrum of $T$ waves at turbulent equilibrium
have already been made in the case of Maxwellian plasmas,
but to the best of our knowledge have not yet been made taking
into account the presence of a tenuous but energetic population
of Kappa distributed particles. In addition to the
investigation of the equilibrium spectra, we also investigate
using weak turbulence theory, the time evolution of the
wave-particle system when an electron beam is assumed to
exist in the medium.

The equations of weak turbulence theory can be found in the
literature and will not be reproduced here, for the sake of economy
of space. For the present paper we utilize the formalism as
presented in Ref. \cite{pl:YoonZGP12}, and only comment on
basic features of these equations, which will be useful for
the analysis of the results appearing in the present paper.
We start by commenting on the equation which describes the time
evolution of $L$ waves.

In the context of weak turbulence theory, the time evolution
of $L$ waves is ruled by terms associated with spontaneous
and induced emission, three-wave decay, and spontaneous plus
induced scattering. The emission terms satisfy the wave-particle
resonance condition,
$\sigma\omega_{\bf k}^L-{\bf k}\cdot{\bf v}=0$, where
$\omega_{\bf k}^L$ is the dispersion relation for $L$ waves,
and $\sigma=\pm 1$ represent forward or backward propagation
of the waves. The three-wave decay processes involve
interactions between different types of waves, satisfying
the following resonance conditions:
$\sigma\omega_{\bf k}^L-\sigma'\omega_{\bf k'}^L
-\sigma''\omega_{\bf k-k'}^S=0$,
$\sigma\omega_{\bf k}^L-\sigma'\omega_{\bf k'}^L
-\sigma''\omega_{\bf k-k'}^T=0$,
$\sigma\omega_{\bf k}^L-\sigma'\omega_{\bf k'}^T
-\sigma''\omega_{\bf k-k'}^T=0$, and
$\sigma\omega_{\bf k}^L-\sigma'\omega_{\bf k'}^S
-\sigma''\omega_{\bf k-k'}^T=0$, where $\omega_{\bf k}^S$ and
$\omega_{\bf k}^T$ are the dispersion relations
for ion-acoustic waves ($S$) and for transverse waves,
respectively. The scattering processes involve waves with two
different wavelengths and frequencies, interacting with plasma
particles, satisfying the following resonance conditions:
$\sigma\omega_{\bf k}^L-\sigma'
\omega_{\bf k'}^L-({\bf k}-{\bf k'})\cdot{\bf v}=0$, and
$\sigma\omega_{\bf k}^L-\sigma'
\omega_{\bf k'}^T-({\bf k}-{\bf k'})\cdot{\bf v}=0$.
Detailed expressions for these terms can be found,
for instance, in Ref.\ \cite{pl:YoonZGP12}.

The equation that describes the time evolution of
$S$ waves presents a similar structure,
containing spontaneous and induced emission terms, which satisfy
the resonance condition,
$\sigma\omega_{\bf k}^S-{\bf k}\cdot{\bf v}=0$;
three-wave decay terms satisfying the resonance conditions,
$\sigma\omega_{\bf k}^S-\sigma'
\omega_{\bf k'}^L-\sigma''\omega_{\bf k-k'}^L=0$ and
$\sigma\omega_{\bf k}^S-\sigma'\omega_{\bf k'}^L
-\sigma''\omega_{\bf k-k'}^T=0$;
and a scattering term which satisfies
$\sigma\omega_{\bf k}^S-\sigma'
\omega_{\bf k'}^L-({\bf k}-{\bf k'})\cdot{\bf v}=0$.
The scattering processes are deemed to be extremely
slow in the case of $S$ waves, and are usually
neglected \cite{pl:YoonZGP12}.

The equation for the $T$ waves can be considered of a
different nature, in the sense that the superluminal $T$
waves do not satisfy the wave-particle resonance condition,
and therefore there is no emission terms, either spontaneous
or induced. The equation which describes the
time evolution of $T$ waves
features three-wave decay terms with resonance conditions given by
$\sigma\omega_{\bf k}^T-\sigma'\omega_{\bf k'}^L
-\sigma''\omega_{\bf k-k'}^L=0$,
$\sigma\omega_{\bf k}^T-\sigma'\omega_{\bf k'}^L
-\sigma''\omega_{\bf k-k'}^S=0$, and
$\sigma\omega_{\bf k}^T-\sigma'\omega_{\bf k'}^T
-\sigma''\omega_{\bf k-k'}^L=0$,
and a scattering term satisfying $\sigma\omega_{\bf k}^T
-\sigma'\omega_{\bf k'}^L-({\bf k}-{\bf k'})\cdot{\bf v}=0$
\cite{pl:YoonZGP12}.

In addition to the wave equations, the set of weak turbulence
equations also contains equations for the time 
evolution of the particle distribution functions. In
collisionless plasmas, the equation for the time evolution of
the particle distribution function is
well-known (see, for instance, equation (1) of 
Ref.\ \cite{pl:YoonZGP12}), and includes 
a quasilinear diffusion term and a term
originated from spontaneous fluctuations, both satisfying
the wave-particle resonance conditions
$\sigma\omega_{\bf k}^\alpha
-{\bf k}\cdot{\bf v}=0$, where $\alpha$ can be $L$ or $S$,
\begin{eqnarray}
\label{fa,eq}
\frac{\partial f_a({\bf v})}{\partial t}
&=& \frac{\pi e_a^2}{m_a^2}\sum_{\sigma=\pm1}
\sum_{\alpha=L,S}\int\frac{d{\bf k}}{k^2}\;
{\bf k}\cdot\frac{\partial}{\partial{\bf v}}\;
\delta(\sigma\omega_{\bf k}^\alpha-{\bf k}\cdot{\bf v})
\nonumber\\
&& \times\,\left(\frac{m_a\sigma
\omega_{\bf k}^\alpha}{4\pi^2}\,f_a({\bf v})
+I_{\bf k}^{\sigma\alpha}\,{\bf k}\cdot
   \frac{\partial f_a({\bf v})}{\partial{\bf v}}\right).
\end{eqnarray}
In equation (\ref{fa,eq}), $f_a({\bf v})$ is the distribution 
function for particles of species $a$ ($a=e$ for electrons and 
$a=i$ for ions), normalized as $\int d{\bf v}\;f_a({\bf v})= 1$.

The present paper is organized as follows:
In section \ref{sec:veloc-distr} we introduce a generic
form of isotropic Kappa distribution, and describe the
distribution function for plasma particles, constituted
by a summation of a Maxwellian distribution and an isotropic
Kappa distribution, with much lower number density than the
Maxwellian population. In section \ref{sec:initial,LS}
we briefly derive the expressions which give the spectra of
$L$ and $S$ waves, and that satisfy equilibrium conditions.
In so doing, we take into account the velocity distributions
presented in section \ref{sec:veloc-distr}.
In section \ref{sec:asymptotic,T} we discuss the possibility
of an asymptotic spectrum of transverse waves ($T$),
which is the result of nonlinear interaction in the
wave-particle system. We derive an expression, which
approximately describes this asymptotic state.
In section \ref{sec:numerical} we present some results
that show the wave spectra, taking into account parameters,
which are compatible with conditions in the solar wind.
We also present in section \ref{sec:numerical} some results
that show the time evolution of the wave spectra and of
the particle distribution function, obtained by numerical
solution of equations of weak turbulence theory.
Section \ref{sec:final}  summarizes the results obtained.

\section{The velocity distributions for plasma particles}
\label{sec:veloc-distr}

Let us assume that isotropic distributions for ions and electrons
are made of the summation of Maxwellian and Kappa distributions.
In three dimensions (3D), considering a generic form for
the Kappa distribution, we may write
\begin{equation}
\label{fbeta}
f_{\beta}({\bf v}) = \left(1-\frac{n_{\kappa\beta}}{n_e}\right)
f_{\beta,M}({\bf v})+\frac{n_{\kappa\beta}}{n_e}
f_{\beta,\kappa}({\bf v}),
\end{equation}
where
\begin{eqnarray} 
\label{fbeta,M}
&f_{\beta,M}({\bf v}) = \frac{1}{\pi^{3/2} v_{\beta}^3}
\exp\left(-\frac{v^2}{v_\beta^2}\right),&\\
\label{fbeta,kappa}
&f_{\beta,\kappa}({\bf v}) = \frac{1}
{\pi^{3/2}\kappa_\beta^{3/2} v_{\beta,\kappa}^3}
\frac{\Gamma(\kappa_\beta+\alpha_\beta)}
{\Gamma\left(\kappa_\beta+\alpha_\beta-{3\over2}\right)} 
\left(1+\frac{v^2}{\kappa_\beta v_{\beta,\kappa}^2}
\right)^{-(\kappa_\beta+\alpha_\beta)},&\nonumber\\
&&\,
\end{eqnarray}
where $v_\beta=\sqrt{2T_\beta/m_\beta}$ is the thermal velocity
of particle species labeled $\beta$, and $v_{\beta,\kappa}$ is
a parameter with the same physical dimension as the particle
thermal velocity, and which reduces to the thermal velocity
in the limit $\kappa_\beta\to\infty$. The distribution functions
given by Eqs.\ (\ref{fbeta,M}) and (\ref{fbeta,kappa}) are
normalized such that $\int d^3v\,f_{\beta}=1$.

Particular cases of the distribution (\ref{fbeta,kappa}) that
correspond to forms of Kappa distributions which are widely
used in the literature can be obtained by a suitable choice
of parameters $\alpha_\beta$ and $v_{\beta,\kappa}$. Namely,  
if $\alpha_\beta=1$, and
\begin{equation}
\label{w2,ST}
v_{\beta,\kappa}^2=\frac{\kappa_\beta-{3\over2}}
{\kappa_\beta}\,v_{\beta}^2,
\end{equation}
then Eq.\ (\ref{fbeta,kappa}) becomes a form of isotropic
Kappa distribution, which is widely used in the
literature \cite{Vasyliunas68, SummersT91, MaceHellberg95},
\begin{equation}
\label{fkappa,ST}
f_{\beta}({\bf v})=\frac{1}{\pi^{3/2}\kappa_\beta^{3/2}
v_{\beta,\kappa}^3}\frac{\Gamma(\kappa_\beta+1)}
{\Gamma\left(\kappa_\beta-{1\over2}\right)}
\left(1+\frac{v^2}{\kappa_\beta v_{\beta,\kappa}^2}
\right)^{-(\kappa_\beta+1)}.
\end{equation}
The average value of the kinetic energy, in the case of
distribution (\ref{fkappa,ST}), leads to the usual notion
of temperature, since it is easily obtained that
\begin{equation}
\left<\frac{1}{2}\,m v^2\right>_\beta=\frac{3T_\beta}{2}.
\end{equation}
In the above, $\left<\cdots\right>_\beta=\int d{\bf v}\cdots f_\beta$.

Another customary choice is to take $\alpha_\beta=0$ and
$v_{\beta,\kappa}=v_{\beta}$. This corresponds to the case
in which Eq.\ (\ref{fbeta,kappa}) becomes the isotropic
Kappa distribution, which is used, for instance, in
\cite{Leubner02, Leubner04b},
\begin{equation}
\label{fkappa,Le}
f_{\beta}({\bf v})=\frac{1}{\pi^{3/2}\kappa_\beta^{3/2}
v_{\beta}^3}\frac{\Gamma(\kappa_\beta)}
{\Gamma\left(\kappa_\beta-{3\over2}\right)}
\left(1+\frac{v^2}{\kappa_\beta v_{\beta}^2}
\right)^{-\kappa_\beta}.
\end{equation}
For distribution function (\ref{fkappa,Le}), the average value
of the kinetic energy does not lead to the usual notion
of temperature, since it is easy to obtain that
\begin{equation}
\left<\frac{1}{2}\,m v^2\right>_\beta 
=\frac{3T_{\beta}}{2}\frac{\kappa_\beta}{\kappa_\beta-5/2}.
\end{equation} 

It can also be noticed that the  distribution given by
(\ref{fkappa,Le}) can be obtained as a result of the use of
the non-extensive statistical mechanics as formulated in
\cite{Tsallis88, SilvaPL98, Leubner02}, while the
distribution function given by (\ref{fkappa,ST}) results
from a modified approach to non-extensive statistical
mechanics, which utilizes the so-called escort probability
functions \cite{TsallisMP98, LivadiotisM09}.

For convenience, we define dimensionless velocities,
by division of the velocity by the electron 
thermal velocity $v_e$, ${\bf u}= {\bf v}/v_e$, the normalized
wavenumber ${\bf q}= {\bf k}v_e/\omega_{pe}$, the 
normalized wave frequency for waves of type $\alpha$,
$z_{\bf q}^\alpha= \omega_{\bf q}^\alpha/\omega_{pe}$ (where
$\alpha= L$, $S$, or $T$), and the dimensionless time variable,
$\tau=\omega_{pe}t$, with $\omega_{pe}=\sqrt{4\pi n_ee^2/m_e}$
being the electron plasma frequency. We also define the normalized
wave intensity for waves of type $\alpha$,
\begin{equation}
{\cal E}_{\bf q}^{\sigma\alpha}= \frac{(2\pi)^2 g}{m_ev_e^2} 
\frac{I_{\bf k}^{\sigma\alpha}}{\mu_{\bf k}^\alpha},
\end{equation}
and introduce other useful dimensionless quantities,
\begin{eqnarray}
\label{norm}
&& u=\frac{v}{v_e},\quad
u_{\beta,\kappa}=\frac{v_{\beta,\kappa}}{v_e},\quad
u_\beta=\frac{v_\beta}{v_e},
\nonumber\\
&& \mu=\frac{m_e}{m_i},\quad
\delta_e=\frac{n_{\kappa e}}{n_e},\quad
\delta_i=\frac{n_{\kappa i}}{n_e}.
\end{eqnarray}

In terms of the dimensionless variables the dispersion relations for the
waves and the velocity distribution
functions become as follows:
\begin{eqnarray}
\label{zL}
	&z_{\bf q}^L= \left(1+\frac{3}{2}q^2\right)^{1/2},&\\
\label{zS}
	&z_{\bf q}^S= \frac{q}{\sqrt{2}} \left(\frac{m_e}{m_i}\right)^{1/2} 
	\left(1+3\frac{T_i}{T_e}\right)^{1/2}
	\left(1+\frac{1}{2}q^2\right)^{-1/2},&\\
\label{zT}
 &z_{\bf q}^T= \left(1+\frac{c^2}{v_e^2}q^2\right)^{1/2},&\\
\label{fbeta,M,2}
&\Phi_{\beta,M}({\bf u}) = \frac{1}{\pi^{3/2} u_{\beta}^3}
\exp\left(-\frac{u^2}{u_\beta^2}\right),&\\
\label{fbeta,kappa,2}
&\Phi_{\beta,\kappa}({\bf u}) = \frac{1}{\pi^{3/2}
\kappa_\beta^{3/2}u_{\beta,\kappa}^3}
\frac{\Gamma(\kappa_\beta+\alpha_\beta)}
{\Gamma\left(\kappa_\beta+\alpha_\beta-{3\over2}\right)}
\left(1+\frac{u^2}{\kappa_\beta u_{\beta,\kappa}^2}
\right)^{-(\kappa_\beta+\alpha_\beta)}.&\nonumber\\
&&\,  
\end{eqnarray}

\section{Initial $L$ and $S$ wave intensities}
\label{sec:initial,LS}

Making use of the equations of weak turbulence theory,
the spectra of electrostatic waves may be initialized by
neglecting the nonlinear interactions and
balancing the spontaneous and induced emission terms, and
by taking into account only the background populations.
For the $L$ waves, using the symbol $\Phi_e({\bf u})$ for
the electron distribution function in terms of
normalized quantities, we utilize the wave equation 
without the nonlinear terms, written in 
terms of the dimensionless quantities,
\begin{eqnarray}
\frac{\partial}{\partial\tau}{\cal E}_{\bf q}^{\sigma L}
=&&\frac{\pi}{q^2}\int d{\bf u}\;\delta(\sigma z_{\bf q}^L
-{\bf q}\cdot{\bf u})\nonumber\\
&&\times\left(g\,\Phi_e({\bf u})+(\sigma z_{\bf q}^L)\,
{\bf q}\cdot\frac{\partial \Phi_e({\bf u})}{\partial{\bf u}}
\,{\cal E}_{\bf q}^{\sigma L}\right).\qquad 
\label{Lwave,eq}
\end{eqnarray}

Using spherical coordinates in velocity space, with the $z$ axis
along ${\bf q}$, and considering distribution (\ref{fbeta})
written in terms of dimensionless variables, we obtain
\begin{eqnarray}
\frac{\partial}{\partial\tau}{\cal E}_{\bf q}^{\sigma L}
&&= \frac{\pi}{q^2}\biggl\{
g\left[\left(1-\delta_e\right)I_M^{eL}
+\delta_e I_1^{eL}\right]\nonumber\\
-2&&(\sigma z_{\bf q}^L)^2
\left[\left(1-\delta_e\right)I_M^{eL}
+\frac{\delta_e u_e^2}{u_{e,\kappa}^2}
\frac{(\kappa_e+\alpha_e)}{\kappa_e}\,I_2^{eL}
   \right]{\cal E}_{\bf q}^{\sigma L}\biggr\},\nonumber\\
 && 
\end{eqnarray}
where
\begin{eqnarray}
I_M^{\beta\alpha} &=& \int d^3u\; \Phi_{\beta,M}(u)
\delta(\sigma z_{\bf q}^\alpha-{\bf q}\cdot{\bf u}),
\nonumber\\
I_1^{\beta\alpha} &=& \int d^3u\; \Phi_{\beta,\kappa}(u)
\delta(\sigma z_{\bf q}^\alpha-{\bf q}\cdot{\bf u}),
\\
I_2^{\beta\alpha} &=& \int d^3u\;
\left(1+\frac{u^2}{\kappa_\beta u_{\beta,\kappa}^2}\right)^{-1}
\Phi_{\beta,\kappa}(u)\delta(\sigma z_{\bf q}^\alpha
-{\bf q}\cdot{\bf u}).
\nonumber
\end{eqnarray}

The equilibrium is obtained by setting the expression for
the time derivative equal to zero, which leads to
\begin{equation}
{\cal E}_{\bf q}^{\sigma L}= \frac{g}{2(z_{\bf q}^L)^2}
\frac{\left(1-\delta_e\right)I_M^{eL}+\delta_e I_1^{eL}}
{\left(1-\delta_e\right)I_M^{eL}
+\frac{\delta_e u_e^2}{u_{e,\kappa}^2}
\frac{(\kappa_e+\alpha_e)}{\kappa_e}\,I_2^{eL}}.
\end{equation}

The integrals $I_M^{\beta\alpha}$, $I_1^{\beta\alpha}$ and
$I_2^{\beta\alpha}$ can be evaluated analytically.
For the case $\beta=e$, it is possible to obtain
\begin{widetext}
\begin{eqnarray}
\label{L,initial,1}
{\cal E}_{\bf q}^{\sigma L} &=& \frac{g}{2(z_{\bf q}^L)^2}
\left[\left(1-\delta_e\right)\exp\left(-\xi_e\right)
+\frac{\delta_e u_e}{\kappa_e^{1/2} u_{e,\kappa}}\nonumber
\frac{\Gamma(\kappa_e+\alpha_e-1)}{\Gamma(\kappa_e+\alpha_e-3/2)}
\frac{1}{\left(1+\xi_{e,\kappa}\right)^{\kappa_e+\alpha_e-1}}\right]
\nonumber\\
&& \times \left[\left(1-\delta_e\right)\exp\left(-\xi_e\right)
+\frac{\delta_e u_e^3}{\kappa_e^{3/2} u_{e,\kappa}^3}
\frac{\Gamma(\kappa_e+\alpha_e)}{\Gamma(\kappa_e+\alpha_e-3/2)}
\frac{1}{\left(1+\xi_{e,\kappa}\right)^{\kappa_e+\alpha_e}}\right]^{-1},
\\
\xi_e &=& \frac{(z_q^L/q)^2}{u_{e}^2},\qquad
\xi_{e,\kappa}=\frac{(z_q^L/q)^2}{\kappa_e u_{e,\kappa}^2}.\nonumber
\end{eqnarray}

For the $S$ waves, we obtain
the following equation
\begin{eqnarray}
\label{Swave,eq}
\frac{\partial}{\partial\tau}{\cal E}_{\bf q}^{\sigma S}
= \mu_{\bf q}^S\,\frac{\pi}{q^2}\int d{\bf u}\;
\delta(\sigma z_{\bf q}^S-{\bf q}\cdot{\bf u})
\left[g\left(\Phi_e({\bf u})+\Phi_i({\bf u})\right)\right.\\
\left. +(\sigma z_{\bf q}^L)\,
\left({\bf q}\cdot\frac{\partial \Phi_e({\bf u})}
{\partial{\bf u}}+\frac{m_e}{m_i}{\bf q}\cdot
\frac{\partial \Phi_i({\bf u})}{\partial{\bf u}}\right)
{\cal E}_{\bf q}^{\sigma S}\right], \nonumber
\end{eqnarray}
\end{widetext}
where
\begin{equation}
\mu_{\bf q}^S= \frac{q^3}{2^{3/2}}
\,\sqrt{\frac{m_e}{m_i}}\left(1+\frac{3T_i}{T_e}\right)^{1/2}.
\end{equation}

Following steps which are similar to those employed in the case
of $L$ waves, the initial spectrum of $S$ waves is seen to obey
the following expression:
\begin{equation}
{\cal E}_{\bf q}^{\sigma S}= \frac{g}{2(z_{\bf q}^L)
(z_{\bf q}^S)}\frac{N_s}{D_s},
\end{equation}
where
\begin{eqnarray}
N_s &=& \left(1-\delta_e\right)I_M^{eS}+\delta_e I_1^{eS}
+\left(1-\delta_i\right)I_M^{iS}+\delta_i I_1^{iS},
\nonumber\\
D_s &=& \left(1-\delta_e\right)I_M^{eS}
+\frac{\delta_e u_e^2}{u_{e,\kappa}^2}
\frac{(\kappa_e+\alpha_e)}{\kappa_e}\,I_2^{eS}\nonumber\\
&&+\mu\left(1-\delta_i\right)\frac{u_e^2}{u_i^2}\,I_M^{iS}
+\frac{\delta_i\mu u_e^2}{u_{i,\kappa}^2}
\frac{(\kappa_i+\alpha_i)}{\kappa_i}\,I_2^{iS}. \nonumber
\end{eqnarray}
After evaluation of the $I_M^{\beta\alpha}$, $I_1^{\beta\alpha}$
and  $I_2^{\beta\alpha}$ integrals, one obtains the following,
\begin{widetext}
\begin{eqnarray}
\label{S,initial,1}
{\cal E}_{\bf q}^{\sigma S} &=& 
\frac{g}{2(z_{\bf q}^L)(z_{\bf q}^S)}
\left[\left(1-\delta_e\right)\exp\left(-\xi_e\right)
+\frac{\delta_e u_e}{\kappa_e^{1/2} u_{e,\kappa}}
\frac{\Gamma(\kappa_e+\alpha_e-1)}{\Gamma(\kappa_e+\alpha_e-3/2)}
\frac{1}{\left(1+\xi_{e,\kappa}\right)^{\kappa_e+\alpha_e-1}}\right.
\nonumber\\
&& \left.+\left(1-\delta_i\right)\exp\left(-\xi_i\right)
+\frac{\delta_i u_e}{\kappa_i^{1/2} u_{i,\kappa}}
\frac{\Gamma(\kappa_i+\alpha_i-1)}{\Gamma(\kappa_i+\alpha_i-3/2)}
\frac{1}{\left(1+\xi_{i,\kappa}\right)^{\kappa_i+\alpha_i-1}}\right]
\nonumber\\
&& \times\left[\left(1-\delta\right)\exp\left(-\xi_e\right)
+\frac{\delta u_e^3}{\kappa_e^{3/2} u_{e,\kappa}^3}
\frac{\Gamma(\kappa_e+\alpha_e)}{\Gamma(\kappa_e+\alpha_e-3/2)}
\frac{1}{\left(1+\xi_{e,\kappa}\right)^{\kappa_e+\alpha_e}}\right.
\nonumber\\
&& \left.+\mu\left(1-\delta_i\right)\frac{u_e^3}{u_i^3}
\exp\left(-\xi_i\right)+\frac{\delta_i\mu u_e^3}
{\kappa_i^{3/2} u_{i,\kappa}^3}
\frac{\Gamma(\kappa_i+\alpha_i)}{\Gamma(\kappa_i+\alpha_i-3/2)}
\frac{1}{\left(1+\xi_i\right)^{\kappa_i+\alpha_i}}\right]^{-1},
\end{eqnarray}
\end{widetext}
where $\xi_e$ and $\xi_{e,\kappa}$ are defined in
(\ref{L,initial,1}) and
\begin{equation}
\xi_i=\frac{(z_q^L/q)^2}{u_{i}^2},\qquad
\xi_{i,\kappa}=\frac{(z_q^L/q)^2}{\kappa_i u_{i,\kappa}^2}.
\end{equation}

\vspace{-0.5cm}
This brings a closure to the first part of the present paper,
namely, to theoretically discuss the self-consistent form
of electrostatic Langmuir and ion-sound wave fluctuation
intensities that arise when the electron velocity distribution
function is composed of a Maxwellian core plus a ``halo''
component given by a Kappa distribution. In Ref.\ \cite{Kim+16}
a similar problem was approached (minus the discussion of
ion-acoustic wave intensity) by considering the iterative
numerical solution of the self-consistent set of particle
and wave kinetic equations. The present discussion complements
Ref.\ \cite{Kim+16} in that our approach has been within the context
of an analytical method. The analytical solution, while less
rigorous that the iterative solution obtain by Ref.\ \cite{Kim+16},
is nevertheless useful in the subsequent discussion of
transverse wave intensity, which we turn to next.

\section{Asymptotic wave level for transverse waves}
\label{sec:asymptotic,T}

The time evolution of transverse $T$ waves, which are
electromagnetic waves, is governed by an equation that
contains terms related to three wave decay involving a
$T$ wave and two $L$ waves, a $T$ wave, a $L$ wave and a
$S$ wave, and two $T$ waves and a $L$ wave, and also a
scattering term involving a $T$ wave, a $L$ wave, and
particles \cite{pl:YoonZGP12}. The evolution equation does
not feature a quasilinear term, like those appearing in the
equations for $L$ and $S$ waves, given by Eqs.\ (\ref{Lwave,eq})
and (\ref{Swave,eq}), because the linear resonance condition
with the particles is not satisfied by the superluminal $T$
waves \cite{pl:YoonZGP12}.

The occurrence of decay processes involving $L$ and $S$ waves,
and also of scattering processes, has as a consequence that $T$
waves are generated by these nonlinear mechanisms, even if they
are not considered present as an initial condition. It is therefore
pertinent to investigate the asymptotic state attained by the
spectrum of $T$ waves, due to the nonlinear processes.
This asymptotic state characterizes what can be called a
``turbulent equilibrium,'' and has already been investigated by
us considering an equilibrium plasma in which the plasma particles
are described by  Maxwellian distributions \cite{pl:ZiebellYSGP14,
pl:ZiebellYGP14}. In the present investigation, we consider the case
in which the velocity distributions of plasma particles contain
a population described by Kappa distributions, as given by Eq.\
(\ref{fbeta}).

At the asymptotic state, it may be considered that the decay
terms are not very effective in changing the wave level, since
they do not involve particles and represent just an exchange
of momentum and energy among different waves. The scattering term
can therefore be considered to be the dominant term for late
evolution of the system. This conjecture has already been used
in the case of Maxwellian velocity distributions, and has been
well supported by numerical analysis of the time evolution
considering the complete weak turbulence equation for the $T$
waves \cite{pl:ZiebellYSGP14, pl:ZiebellYGP14}. Consequently,
using this approximation and adopting normalized variables,
the equation for late stages of the time evolution of $T$ waves
can be written as follows \cite{pl:YoonZGP12},
\begin{widetext}
\begin{eqnarray}
\frac{\partial}{\partial\tau}{\cal E}_{\bf q}^{\sigma T}
&\simeq& \sum_{\sigma'}\int d{\bf q}'\int d{\bf u}\;
\frac{({\bf q}\times{\bf q}')^2}{q^2\,q'^2}\;
\delta\left[\sigma z_{\bf q}^T
-\sigma'z_{{\bf q}'}^L-({\bf q}-{\bf q}')\cdot{\bf u}\right]
\nonumber\\
&& \times\left[g(\sigma z_{\bf q}^T)
\left(\sigma z_{\bf q}^T
\,{\cal E}_{{\bf q}'}^{\sigma'L}
-\sigma'z_{{\bf q}'}^L
\,\frac{{\cal E}_{\bf q}^{\sigma T}}{2}\right)
\left(\Phi_e+\Phi_i\right)\right.
\nonumber\\
&& \left.-\,{\cal E}_{{\bf q}'}^{\sigma'L}
\frac{{\cal E}_{\bf q}^{\sigma T}}{2}\;({\bf q}-{\bf q}')
\cdot\frac{\partial}{\partial{\bf u}}
\left((\sigma z_{\bf q}^T-\sigma'z_{\bf q'}^L)\Phi_e -\frac{m_e}{m_i}
(\sigma z_{\bf q}^T)\Phi_i\right)\right].
\end{eqnarray}

The asymptotic state is obtained by taking the time derivative
equal to zero. Doing this, and using the distribution functions
given by Eq.\ (\ref{fbeta}), we obtain
\begin{eqnarray}
\sum_{\sigma'}\int d{\bf q}'\;&&
\frac{({\bf q}\times{\bf q}')^2}{q^2\,q'^2}
\left\{g(\sigma z_{\bf q}^T)\left(
\sigma'z_{{\bf q}'}^L\,\frac{{\cal E}_{\bf q}^{\sigma T}}{2}
-\sigma z_{\bf q}^T \,{\cal E}_{{\bf q}'}^{\sigma'L}\right)
\left[\left(1-\delta_e\right)I_M^e+\delta_e\,I_1^e
+\left(1-\delta_i\right)I_M^i
+\delta_i I_1^i\right]
\nonumber \right.\\
&&+\frac{m_e}{m_i}\,{\cal E}_{{\bf q}'}^{\sigma'L}
\frac{{\cal E}_{\bf q}^{\sigma T}}{2}
(\sigma z_{\bf q}^T-\sigma'z_{{\bf q}'}^L)(\sigma z_{\bf q}^T)
\left[\frac{2}{u_i^2}\left(1-\delta_i\right)I_M^i
+\delta_i\frac{2}{u_{i,\kappa}^2}
\frac{(\kappa_i+\alpha_i)}{\kappa_i}I_2^i\right]\\
&&\left.-\,{\cal E}_{{\bf q}'}^{\sigma'L}
\frac{{\cal E}_{\bf q}^{\sigma T}}{2}
\left(\sigma z_{\bf q}^T-\sigma'z_{{\bf q}'}^L\right)^2
\left[\frac{2}{u_e^2}\left(1-\delta_e\right)I_M^e
+\delta_e\,\frac{2}{u_{e,\kappa}^2}
\frac{(\kappa_e+\alpha_e)}{\kappa_e}I_2^e\right]\right\} = 0. \nonumber
\end{eqnarray}

Making use of the analytical expressions for the quantities
$I_M^\beta$, $I_1^\beta$, and $I_2^\beta$, one arrives at
the following:
\begin{eqnarray}
{\cal E}_{\bf q}^{\sigma T} &=& 2(\sigma z_{\bf q}^T)^2
\sum_{\sigma'}\int d{\bf q}'\;
\frac{({\bf q}\times{\bf q}')^2}{q'^2|{\bf q}-{\bf q'}|}
\,g{\cal E}_{{\bf q}'}^{\sigma'L}\sum_{\beta=e,i}
\left(\frac{1-\delta_\beta}{u_{\beta}}\,e^{-\zeta_\beta}
+\frac{\delta_\beta}{\kappa_\beta^{1/2} u_{\beta,\kappa}}
\frac{\Gamma(\kappa_\beta+\alpha_\beta-1)}
{\Gamma(\kappa_\beta+\alpha_\beta-3/2)}
\frac{1}{\left(1+\zeta_\beta
\right)^{\kappa_\beta+\alpha_\beta-1}}\right)
\nonumber\\
&\times&\left\{\sum_{\sigma'}\int d{\bf q}'\;
\frac{({\bf q}\times{\bf q}')^2}{q'^2|{\bf q}-{\bf q'}|}
\left[g_*(\sigma z_{\bf q}^T)(\sigma'z_{{\bf q}'}^L)
\sum_{\beta=e,i}\left(\frac{1-\delta_\beta}{u_{\beta}}
\,e^{-\zeta_\beta}+\frac{\delta_\beta}{\kappa_\beta^{1/2} u_{\beta,\kappa}}
\frac{\Gamma(\kappa_\beta+\alpha_\beta-1)}
{\Gamma(\kappa_\beta+\alpha_\beta-3/2)}
\frac{1}{\left(1+\zeta_{\beta,\kappa}
\right)^{\kappa_\beta+\alpha_\beta-1}}\right)
\right.\right.\nonumber\\
&+&\mu\,{\cal E}_{{\bf q}'}^{\sigma'L}
(\sigma z_{\bf q}^T-\sigma'z_{{\bf q}'}^L)(\sigma z_{\bf q}^T)
\left(\frac{2\left(1-\delta_i\right)}{u_{i}^3}\,e^{-\zeta_i}
+\frac{2\delta_i}{\kappa_i^{3/2} u_{i,\kappa}^3}
\frac{\Gamma(\kappa_i+\alpha_i)}{\Gamma(\kappa_i+\alpha_i-3/2)}
\frac{1}{\left(1+\zeta_i\right)^{\kappa_i+\alpha_i}}\right)\\
&-&\left.\left.{\cal E}_{{\bf q}'}^{\sigma'L}
(\sigma z_{\bf q}^T-\sigma'z_{{\bf q}'}^L)^2
\left(\frac{2\left(1-\delta_e\right)}{u_{e}^3}\,e^{-\zeta_e}
+\frac{2\delta_e}{\kappa_e^{3/2} u_{e,\kappa}^3}
\frac{\Gamma(\kappa_e+\alpha_e)}{\Gamma(\kappa_e+\alpha_e-3/2)}
\frac{1}{\left(1+\zeta_{e,\kappa}\right)^{\kappa_e+\alpha_e}}
\right)\right]\right\}^{-1},
\nonumber\\
\zeta_\beta &=& \frac{(\sigma z_{\bf q}^T-\sigma'z_{\bf q'}^L)^2}
{u_{\beta}^2|{\bf q}-{\bf q'}|^2},\qquad
\zeta_{\beta,\kappa}=\frac{(\sigma z_{\bf q}^T
-\sigma'z_{\bf q'}^L)^2}{\kappa_\beta u_{\beta,\kappa}^2
|{\bf q}-{\bf q'}|^2},\qquad(\beta=e,i).
\nonumber
\end{eqnarray}

This is a fairly complex expression. However, one notices that the
contributions due to the Maxwellian population feature an exponential
factor, which is very peaked with maximum occurring for
$\sigma'=\sigma$ and $q'\simeq q_m$, the value of $q$ for which
$z_{q'}^L= z_q^T$. The contributions due to the Kappa distribution
are also proportional to a  factor which is unity for
$\sigma'=\sigma$ and $q'\simeq q_m$, and decrease rapidly away
from this point. As a consequence, the terms corresponding to
the induced scattering can be neglected, since they are
proportional to $(\sigma z_{\bf q}^T-\sigma'z_{\bf q'}^L)$,
and the asymptotic spectrum of $T$ waves can be approximated
by the following:\\
\begin{eqnarray}
{\cal E}_{\bf q}^{\sigma T} &\simeq& 2(\sigma z_{\bf q}^T)^2 
\sum_{\sigma'}\int d{\bf q}'\;\frac{({\bf q}\times{\bf q}')^2}
{q'^2|{\bf q}-{\bf q'}|}\,g{\cal E}_{{\bf q_m}}^{\sigma'L}
\sum_{\beta=e,i}\left(\frac{1-\delta_\beta}{u_{\beta}}
\,e^{-\zeta_\beta}+\frac{\delta_\beta}
{\kappa_\beta^{1/2} u_{\beta,\kappa}}
\frac{\Gamma(\kappa_\beta+\alpha_\beta-1)}
{\Gamma(\kappa_\beta+\alpha_\beta-3/2)}
\frac{1}{\left(1+\zeta_{\beta,\kappa}
\right)^{\kappa_\beta+\alpha_\beta-1}}\right)
\nonumber\\
&\times&\left[\sum_{\sigma'}\int d{\bf q}'\;
\frac{({\bf q}\times{\bf q}')^2}{q'^2|{\bf q}-{\bf q'}|}
\,g_*(\sigma z_{\bf q}^T)^2\sum_{\beta=e,i}
\left(\frac{1-\delta_\beta}{u_{\beta}}\,e^{-\zeta_\beta}
+\frac{\delta_\beta}{\kappa_\beta^{1/2} u_{\beta,\kappa}}
\frac{\Gamma(\kappa_\beta+\alpha_\beta-1)}
     {\Gamma(\kappa_\beta+\alpha_\beta-3/2)}
\frac{1}{\left(1+\zeta_{\beta,\kappa}\right)^{\kappa_\beta+\alpha_\beta-1}}
\right)\right]^{-1}.\nonumber\\
 &&
\end{eqnarray}
\end{widetext}

As a result of these approximations, and assuming that in the
absence of  particle beams the spectrum of $L$ waves remains
nearly the same as in the initial state, and is therefore
symmetrical, ${\cal E}_{q_m}^{-L}={\cal E}_{\bf q_m}^{+L}$,
it is seen that the asymptotic spectrum of $T$ waves can be
simplified by
\begin{equation}
\label{T,asympt,1}
{\cal E}_{\bf q}^{\sigma T} 
\simeq 2 {\cal E}_{{\bf q}_m}^{\sigma L}, \quad
q_m= \sqrt{\frac{2}{3}}\frac{c}{v_e}q. 
\end{equation}
To sum up the second part of the present analysis, by making use
of the equations of electromagnetic weak turbulence, we have
derived the asymptotic form of the transverse wave intensity,
which is given in terms of the Langmuir wave intensity.
The Langmuir wave fluctuation intensity, however, was
already discussed in Sec. \ref{sec:initial,LS}, so that we
may readily obtain the explicit form of the transverse
wave intensity.

\section{Numerical results}
\label{sec:numerical}

In order to illustrate the effects of the presence of a Kappa
population of electrons on the spectrum of waves, which satisfy
conditions of equilibrium with the particle distribution, we
consider that the electron population is described by
distribution function (\ref{fbeta}), with $\alpha=1$, and
$u_{e,\kappa}^2= u_e^2(\kappa_e-3/2)/\kappa_e$.
The ion distribution is assumed to be described by an
isotropic Maxwellian distribution, i.e., we assume
$\delta_i=0$. 
Finally, we use $T_e/T_i=2$, a value for 
the electron and ion temperature ratio which is within the
range of values observed in the solar wind 
\cite{Newbury1998}.

\begin{widetext}

  \begin{figure}
    \centering
    \includegraphics[width=0.95\columnwidth]{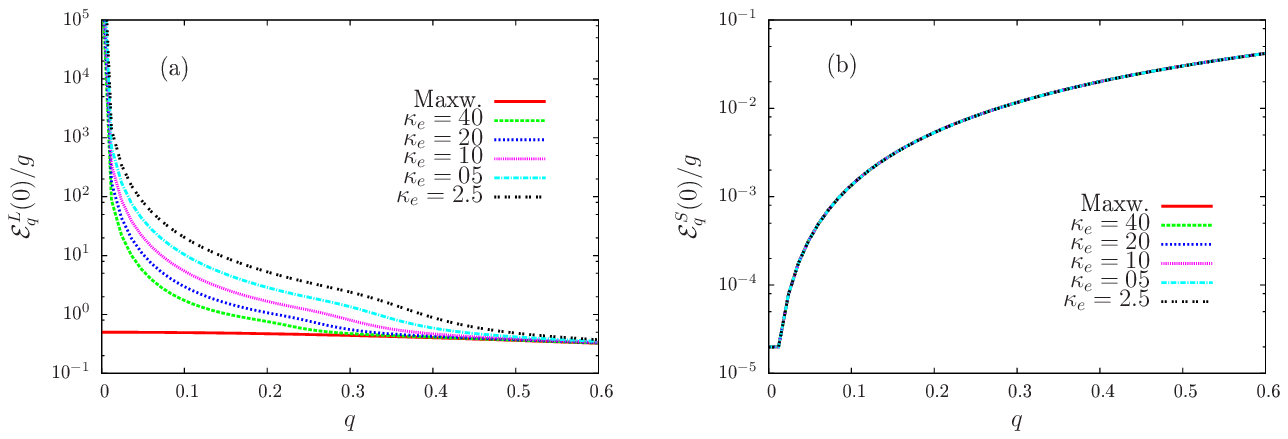}
  \caption{
Initial spectrum of electrostatic waves divided by $g$,
as a function of  normalized wavenumber $q= kv_e/\omega_{pe}$,
for several values of the index $\kappa_e$. The case of
Maxwellian distribution, $n_{\kappa e}/n_e=0.0$, is
also shown for reference. (a) $L$ waves; (b) $S$ waves.
For $S$ waves all curves overlap. Electron distribution given
by Eq.\ \protect{(\ref{fbeta})}, with $\alpha_e=1$ and
$u_{\beta,\kappa}^2= u_\beta^2(\kappa_e-3/2)/\kappa_e$, for
$n_{\kappa e}/n_e=0.1$. The ion distribution is an isotropic
Maxwellian, and $T_e/T_i=2$. The spectra of $L$ and $S$ waves
are given by Eqs.\ \protect{(\ref{L,initial,1})} and
\protect{(\ref{S,initial,1})}.
}\label{fig1}
\centering
\includegraphics[width=0.95\columnwidth]{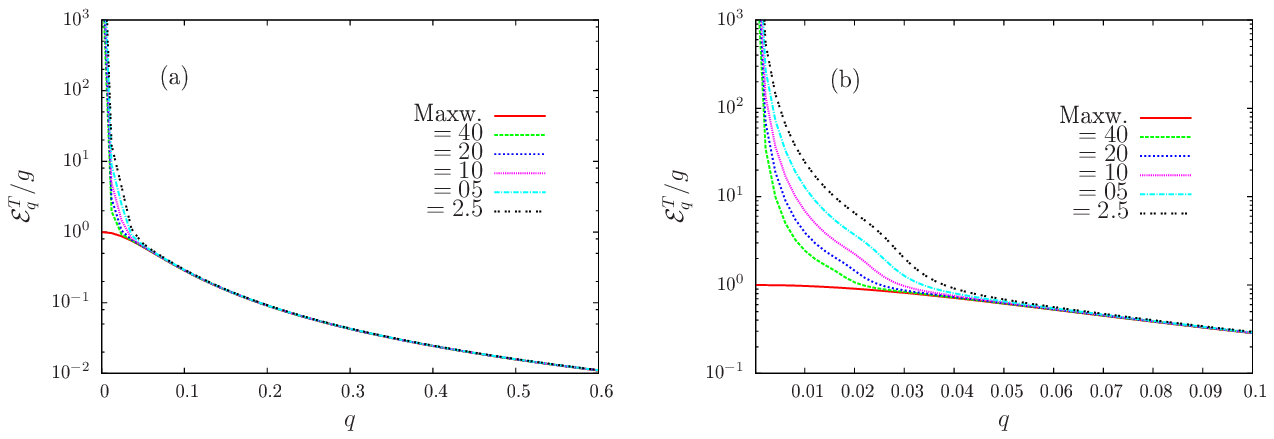}
\caption{
Asymptotic spectrum of $T$ waves, ${\cal E}_{\bf q}^T$,
characterizing a state of ``turbulent
equilibrium'', vs normalized wavenumber $q$.
(a) ${\cal E}_{\bf q}^T$, for $n_{\kappa e}/n_e=0.1$, and several values of
$\kappa_e$. The case of Maxwellian distribution, $n_{\kappa e}/n_e=0.0$, is
also shown for reference;
(b) expanded view of the region of small values of $q$
in \protect{\autoref{fig3}(}a).
Other parameters and conditions are as in \protect{\autoref{fig1}}.
}\label{fig2}
    \centering
\includegraphics[width=0.95\columnwidth]{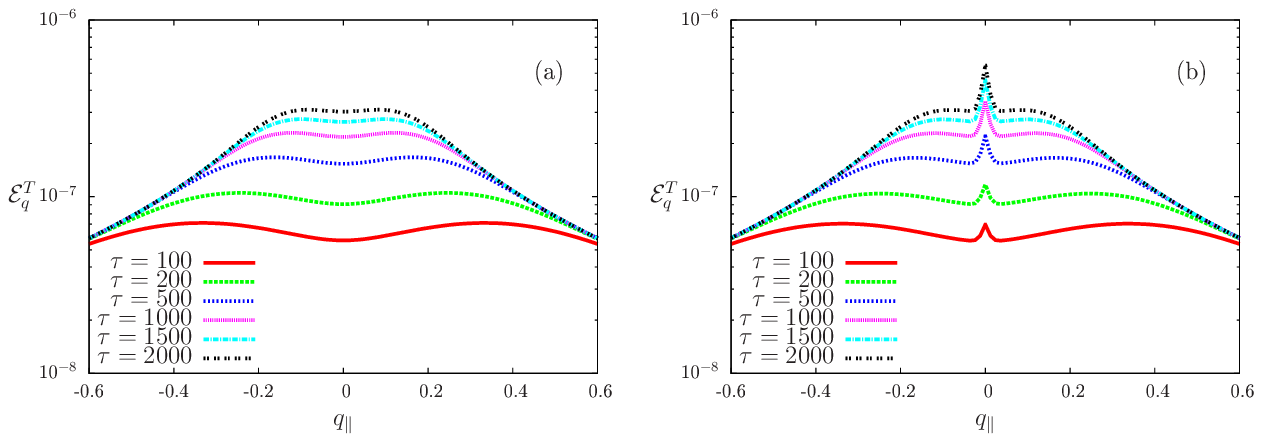}
\caption{
(a) 1D spectrum of $T$ waves vs. $q_\parallel$, for several values of $\tau$,
for $n_{\kappa e}/n_e= 0.0$;
(b) 1D spectrum of $T$ waves vs. $q_\parallel$, for several values of $\tau$,
for $n_{\kappa e}/n_e= 5.0\times 10^{-2}$ and $\kappa_e=5.0$.
Other parameters and conditions are as in \protect{\autoref{fig1}}.
}\label{fig3}
\end{figure}
\end{widetext}

In \autoref{fig1} we show the initial spectrum of electrostatic
waves divided by $g$, as a function of  normalized wave-number
$q= kv_e/\omega_{pe}$, for several values of the index $\kappa_e$.
We assume that the electron population described by a Kappa
distribution constitute $10\%$ of the electron population, i.e.,
$\delta_e=0.1$.
\autoref{fig1}(a) and \autoref{fig1}(b) show the initial
spectra of $L$ and $S$ waves, obtained using Eqs.\
(\ref{L,initial,1}) and (\ref{S,initial,1}), respectively.
The spectra obtained in the case of purely Maxwellian distribution,
with $n_{\kappa e}/n_e=0.0$, are also shown in Figs.\
\ref{fig1}(a) and \ref{fig1}(b), for reference. In Figs.\
\ref{fig1}(a) and \ref{fig1}(b) are shown the curves corresponding
to several values of $\kappa_e$ ($\kappa_e=40$, 20, 10, 5, and 2.5).

\autoref{fig1}(a) shows that the value of ${\cal E}_{\bf q}^L(0)$
in the case of the presence of Kappa population is higher than
the value obtained in the case of a purely Maxwellian distribution,
with a  difference which is already noticeable in the scale of the
figure even for the upper limit shown, $q=0.6$, and increases
for smaller values of $q$, featuring a peak which diverges for
$q\to 0$. For larger values of $\kappa_e$, the shape of the
$L$ spectrum is similar to the shape exhibited
in the case of small values of $\kappa_e$,
but the magnitude of the spectrum at
a given value of $q$ is smaller for increasing values of
$\kappa_e$. It is noticed, however, that the peak at $q\to 0$
is present even for large values of $\kappa_e$.

The presence of the peak in the $L$ spectrum, for $q\to 0$, can be understood
by analysis of Eq. (\ref{L,initial,1}). In the presence of a population of
kappa electrons, even for small value of $n_{\kappa e}/n_e$, it is seen that
for sufficiently small value of $q$ the contribution due to the Maxwellian
population vanishes, due to the factor $\exp\left(-(z_q^L/q)^2/u_e^2\right)$.
For the region of $q$ values where this occurs, the contribution of the
Kappa population is dominant, and the equilibrium spectrum can be given by
the approximated expression,
\begin{equation}
\label{L,initial,2}
{\cal E}_{\bf q}^{\sigma L}\simeq \frac{g}{2(z_{\bf q}^L)^2}
\frac{u_{e,\kappa}^2}{u_e^2}\left(1+\frac{(z_q^L/q)^2}{\kappa_eu_{e,\kappa_e}^2}
\right)
\end{equation}

In the case of $\kappa_e\to \infty$, this expression reduces to
$g/(2(z_{\bf q}^L)^2$, which is the expression obtained in the Maxwellian
case, as expected. However, for finite values of $\kappa_e$, no matter how
large, Eq. (\ref{L,initial,2}) is seen to diverge at $q\to 0$.
The explanation for this is as follows: For large values of $\kappa_e$, the
Kappa distribution coincides with a Maxwellian distribution, in the region
of velocity space with significant electron population. However, the
initial spectra of waves is obtained from Eq. (\ref{Lwave,eq}), which for
equilibrium requires balance between the term associated to spontaneous
fluctuations, which is proportional to the distribution function, and the term
associated to induced emission, which is proportional to the velocity
derivative of the distribution function and to the value of the wave
spectra at the resonant velocity. For $q\to 0$, the resonant velocity
becomes progressively larger. Since for very large velocities the derivative
of the Kappa distribution is smaller than the derivative of the Maxwellian
distribution, the wave spectra for small $q$ has to be higher in the case
of Kappa distribution than in the case of Maxwellian distribution, in
order to satisfy the equilibrium condition.

\autoref{fig1}(b) shows the values of ${\cal E}_{\bf q}^S(0)/g$ vs.
$q=kv_e/\omega_{pe}$. In fact, the figure shows the values of
${\cal E}_{\bf q}^S(0)$ multiplied by $\mu_{\bf q}^S$, but we continue to
denote the quantity as ${\cal E}_{\bf q}^S(0)$, for simplicity. The figure
displays the results obtained for several values of $\kappa_e$, but the
different curves can not be distinguished in the scale of the figure.
It is seen that the kappa index of the Kappa distribution is not relevant for
the initial spectrum of $S$ waves, while it was seen to be relevant for
the initial spectrum of $L$ waves.

We have also obtained the initial spectrum of electrostatic 
waves by assuming fixed value of the index $\kappa_e$ and considering different
values of the relative number density of the Kappa population,
$n_{\kappa e}/n_e$. The results obtained, both for large and small values of 
$\kappa_e$, show that the spectra obtained for $L$ and $S$ waves are almost
independent of the value of the number density of the Kappa population, as 
long it is not zero. These results are not shown here for the sake of 
economy of space, since the curves obtained for different values of
$n_{\kappa e}/n_e$ are basically the same as the curves shown in 
\autoref{fig1}, for each value of $\kappa_e$. The important point to be 
emphasized is that the
presence of a small population of electrons described by a Kappa
population is sufficient to affect significantly the
equilibrium spectrum of $L$ waves in the region of small wave numbers, leading
to the formation of the peaked feature at $q\simeq 0$.

\autoref{fig2} is dedicated to display the asymptotic spectrum of $T$ waves,
obtained using Eq. (\ref{T,asympt,1}). \autoref{fig2}(a) shows
${\cal E}_{\bf q}^T/g$ as a function of normalized wavenumber, for
$n_{\kappa_e}/n_e=0.1$, and several values of $\kappa_e$, and also present
a curve obtained considering a purely Maxwellian electron distribution,
obtained with
$n_{\kappa e}/n_e=0$. The conditions and parameters are the same used to
obtain the spectrum of $L$ waves in \autoref{fig1}. Let us first comment
on the result obtained considering $n_{\kappa_e}/n_e=0$, given by the red
line in \autoref{fig2}(a). This result is explained by analysis of
Eq (\ref{T,asympt,1}), which shows that the spectrum of $T$ waves is
proportional to the spectrum of $L$ waves, given by Eq. (\ref{L,initial,1}),
evaluated at $q=q_m$. If the Kappa population is vanishing,
$n_{\kappa_e}/n_e=0$, the Kappa contributions vanishes in Eq.
(\ref{L,initial,1}), and the contributions due to the Maxwellian population
in the numerator and in the denominator cancel out, and the spectrum turns
out to be given by
\begin{displaymath}
{\cal E}_{\bf q}^{\sigma T} 
\simeq 2 \frac{g}{2(z_{{\bf q}_m}^L)^2}=\frac{g}{2+3q_m^2}.
\end{displaymath}

At $q=0$, the amplitude of the spectrum of $T$ waves in the case of Maxwellian
electron distribution is therefore twice the magnitude of the spectrum of $L$
waves, but decays faster for larger values of $q$, since $q_m>>q$.
With the presence of a population described by a Kappa distribution,
\autoref{fig2}(a) shows that the spectrum of $T$ waves is modified in
the region of small wave numbers, in comparison with the spectrum obtained
in the Maxwellian case. In the scale of the figure, the modification is
noticeable for normalized wave-number $q<0.1$, with a difference that
increases with the decrease of the $\kappa_e$ index, i.e., increases
with the increase of the non-thermal character of the electron distribution.
The spectrum features divergent behavior for
$q\to 0$, as already noticed for the $L$ waves in \autoref{fig1}.

\autoref{fig2}(b) shows an expanded view of the region of small values of 
$q$, for the conditions which have been discussed in \autoref{fig2}(a). The
expanded view clearly shows the increase of the magnitude of the $T$ wave
spectrum at small values of $q$. For instance, it is seen that for $q\simeq 
0.02$ the intensity of the spectrum of $T$ waves in the case of $\kappa_e
=2.5$ is about one order of magnitude above the intensity displayed in the
case of $\kappa_e=40$.

We have also investigated the dependence of the $T$ spectrum on the 
relative number density $n_{\kappa e}/n_e$, for a fixed value of
$\kappa_e$. The results obtained have shown that the $T$ wave
spectrum obtained in the presence of a Kappa distribution is almost independent
of the number density of the Kappa population. The only noticeable feature in
the spectra is the presence of the peak around $q=0$, which occurs for any 
finite value of $n_{\kappa e}/n_e$, and vanishes in the purely Maxwellian case
($n_{\kappa e}/n_e=0$).

In addition to these results concerning the initial spectra of electrostatic
waves and the asymptotic spectrum of transverse waves, we also present some
results which show the time evolution of the wave-particle system, comparing
a situation in which the background
electron velocity distribution is a Maxwellian
distribution with a situation in which an ``halo'' population described by
as isotropic Kappa distribution is also present.

For the study of the time evolution of the system, we
utilize the set of weak turbulence equations, with some additional
approximations. Regarding to the plasma particles, we assume that the
ion velocity distribution remains constant along the evolution, and that
in the case of the equation for the
electron distribution the quasilinear diffusion
due to $S$ waves can be neglected in
comparison with the diffusion caused by the $L$ waves. Regarding the waves,
we describe the evolution of the $L$ waves by including the spontaneous and
induced emission processes, the three-wave decay processes involving
$L$ and $S$ waves, and the scattering process
involving two $L$ waves and the particles.
Nonlinear interaction involving $T$ waves are neglected in the equation for
the time evolution of the $L$ waves, for simplicity, which is common practice
in the literature. The evolution of $S$ waves is described in the present
analysis by taking into account the emission terms and the three-wave decay
term involving $L$ and $S$ waves, and neglecting the effect of the decay
term involving $S$, $L$ and $T$ waves, and also the scattering term. In the
equation for the $T$ waves, however, which contains only nonlinear effects,
we keep all the terms which have already been described in the
introduction section, namely, the decay involving a $T$ wave and two $L$ waves,
the decay involving a $T$ wave, a $L$ wave and a $S$ wave, the decay involving
two $T$ waves and a $L$ wave, and the scattering term.

We utilize a two-dimensional approximation (2D), considering a grid of
51$\times$ 101 points in ($u_\perp,u_z$) space, with $0\le u_\perp\le 12$
and $-12\le u_z\le 12$, a grid of $51\times 51$ points in $(q_\perp,q_z)$
space for $L$ and $S$ waves, and a grid of $71\times 71$ points in
$(q_\perp,q_z$) space for the $T$ waves, which develop fine features which
require better resolution than the $L$ and $S$ waves, considering for all
waves the evolution in the interval $0\le q_\perp\le 0.6$ and
$0\le q_z\le 0.6$. The normalized time step has been adopted as $\Delta\tau
=0.1$, and the equations were solved using a fourth-order Runge-Kutta
procedure for the wave equations and the splitting method for the
equation describing the time evolution of the electrons.

As starting conditions, we assume that the background
electron population is described by distribution function (\ref{fbeta}),
with the 2D versions of Eqs. (\ref{fbeta,M}) and
(\ref{fbeta,kappa}), with
the Kappa distribution defined using $\alpha=1$ and
$u_{e,\kappa}^2= u_e^2(\kappa_e-1)/\kappa_e$, which is the proper value
of $u_{e,\kappa}^2$ for 2D distributions. We assume that the ion
distribution is described by an isotropic Maxwellian distribution,
with $T_e/T_i=2$. We also assume that the plasma parameter is
$(n\lambda_D^3)^{-1}=5.0\times 10^{-3}$, and $v_e^2/c^2= 4.0\times 10^{-3}$,
values which have already been used in analyses of the plasma emission
without taking into account the presence of a Kappa distribution
\cite{pl:ZiebellYGP14,pl:ZiebellYGP14b}.

A further approximation is made for the numerical analysis, regarding the
initial wave spectra. As already discussed in the initial paragraphs of this
section on numerical results (see equation (\ref{L,initial,2}) and the
accompanying comments), in the presence of a Kappa distribution the initial
spectrum of $L$ diverges for $q\to 0$. This divergence, although consistent
with the non-relativistic approach, is not appropriate for the numerical
analysis. In our numerical implementation of the formalism, the initial
spectrum of $L$ waves is given by equation (\ref{L,initial,1}) down to the
value of $q$ such the resonant velocity becomes equal to $c$, i.e., the
value of $q$ for which $z_q^L/q= c/v_e$. It is assumed
that for values of $q$ smaller than this value,
the initial $L$ wave spectrum is given by the
same value obtained at the limit value of the resonant $q$. With such
approximation, when a Kappa distribution is assumed to be present,
the initial spectrum of $L$ has significant growth in the
region of small values of $q$,
in comparison with the spectrum in the case of a Maxwellian plasma
background, but the divergence is avoided. The initial spectrum of $L$ waves
is therefore given by equation (\ref{L,initial,1}), with an approximation in
the region of small values of $q$, and the spectrum of $S$ is given by equation
(\ref{S,initial,1}). The $T$ waves are assumed not present at initial time.

\autoref{fig3} shows one dimensional (1D) representations of the spectrum
of $T$ waves, i.e., obtained after integration of ${\cal E}_{\bf q}^T$
along the perpendicular component of normalized wavenumber, $q_\perp$.
The spectra are shown for different values of $\tau$, and show the evolution
of the $T$ wave spectrum. \autoref{fig3}(a) displays the wave spectra
obtained in the case of purely Maxwellian electron distribution, i.e.,
$n_{\kappa e}/n_e=0.0$. \autoref{fig3}(b) depicts the wave spectra
obtained when the electron distribution contains a Kappa population, with
$n_{\kappa e}/n_e= 5.0\times 10^{-2}$, and $\kappa_e=5.0$. In panel (a), it
is seem that the amplitude of the waves increases for all values of $q_z$,
and gradually evolves toward the asymptotic solution described by equation
(\ref{T,asympt,1}) in the case of $n_{\kappa e}=0.0$, and which appears as
the red lines in \autoref{fig2}. The situation depicted
in \autoref{fig3}(a) corresponds to the initial stages of the evolution,
which is displayed up to longer time in Figure 2 of
Ref. \cite{pl:ZiebellYGP14}. In the presence of a small Kappa population,
the spectrum of $T$ waves evolves as shown in \autoref{fig3}(b). The
spectrum grows for all values of $q_z$, much as seen in \autoref{fig3}(a),
but there is difference. A peak is seen to appear near $q_z=0$, and grows in
time. What is seen is \autoref{fig3}(b) are some steps in the time evolution
of the spectrum which is asymptotically given by equation (\ref{T,asympt,1}),
and represented in \autoref{fig2}. It must be noticed that the peak near
$q_z=0$ in \autoref{fig3}(b) has a finite height. It does not diverge as the
peaks appearing in \autoref{fig2}, because for the numerical analysis of the
equation which describe the time evolution we have assumed that the $L$ wave
spectrum saturates for sufficiently small value of $q$, instead of growing
infinitely for $q\to 0$.

The growth of the peak near $q=0$ in the $T$ wave spectrum can be explained
as follows. The dominant process for the formation of background spectrum of
$T$ is the scattering involving $L$ waves. The scattering effect is maximum
for wave lengths which satisfy $z_q^T=z_{q'}^L$, which means $q'=q_m$,
where $q_m$ is defined by equation (\ref{T,asympt,1}). As seen in
\autoref{fig1}(a), in the case of a small population described by a Kappa
distribution the spectrum of $L$ waves is above the spectra obtained in the
purely Maxwellian case for $q'\le 0.2$. The scattering process is therefore
most effective to generate $T$ waves with $q\le 0.02$, for the value
of $v_e/c$ which we have assumed. The scattering of $L$ waves
is the main responsible
for the formation of the spectrum of $T$ waves, and the peak for large
wavelengths which is seen in the $L$ wave spectrum which occurs in the presence
of Kappa distributed electrons is the cause for the
growth of the peak for large wavelength in the $T$ wave spectrum.

\begin{figure*}
\includegraphics[width=0.9\textwidth]{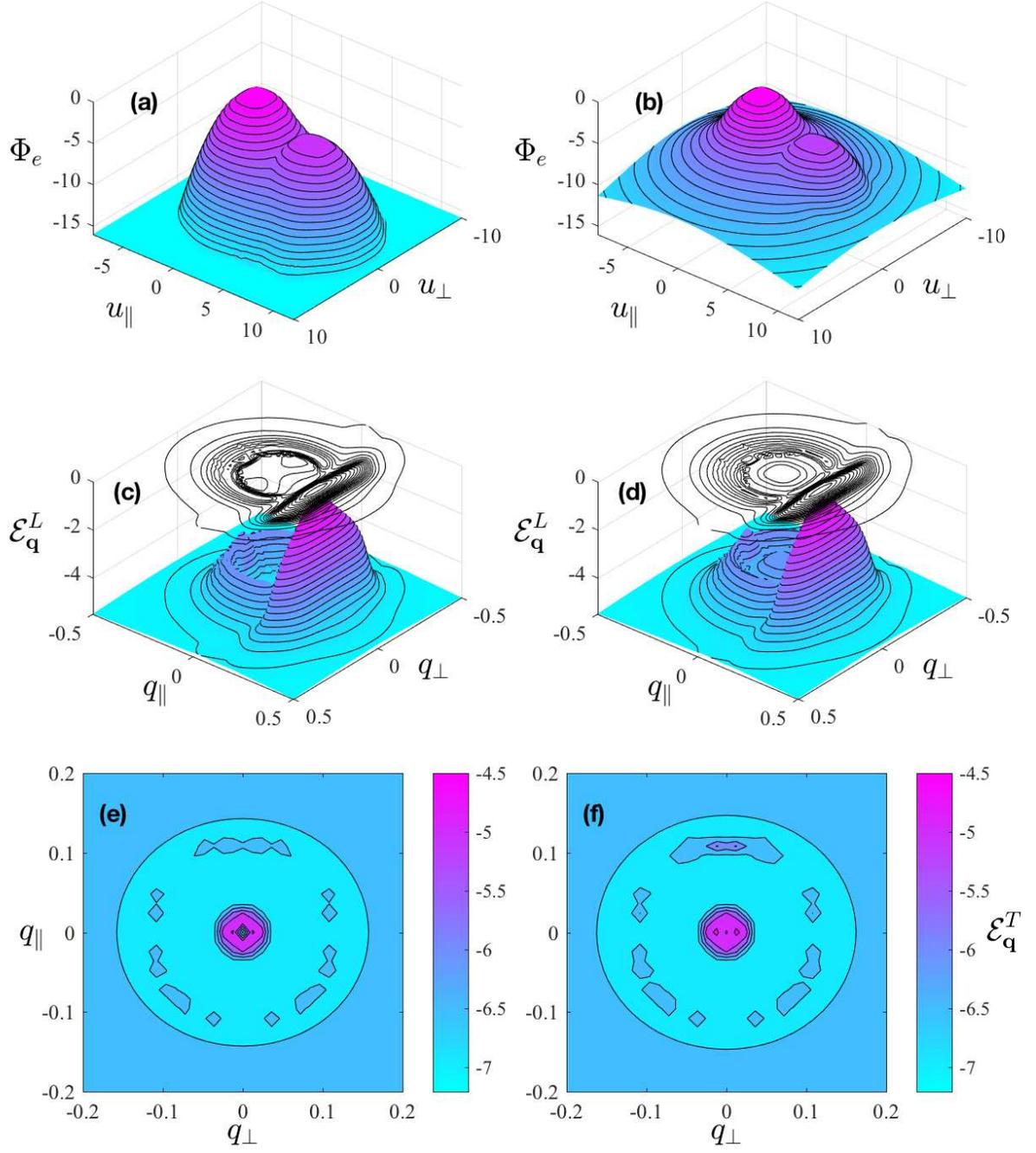}
\caption{
(a) Electron distribution function at $\tau=500$,
vs. $u_\parallel$ and  $u_\perp$, for
$n_{\kappa e}/n_e= 0.0$;
(b) Electron distribution function at $\tau=500$,
vs. $u_\parallel$ and  $u_\perp$, for
$n_{\kappa e}/n_e= 5.0\times 10^{-2}$;
(c) Spectrum of $L$ waves at $\tau=500$,
vs. $q_\parallel$ and  $q_\perp$, for
$n_{\kappa e}/n_e= 0.0$;
(d) Spectrum of $L$ waves at $\tau=500$,
vs. $q_\parallel$ and  $q_\perp$, for
$n_{\kappa e}/n_e= 5.0\times 10^{-2}$;
(e) Spectrum of $T$ waves at $\tau=500$,
vs. $q_\parallel$ and  $q_\perp$,
for $n_{\kappa e}/n_e= 0.0$;
(f) Spectrum of $T$ waves at $\tau=500$,
vs. $q_\parallel$ and  $q_\perp$, for
$n_{\kappa e}/n_e= 5.0\times 10^{-2}$.
Input parameters are as follows: $T_e/T_i= 2.0$,
$T_b/T_e= 1.0$, $n_b/n_e=1.0\times 10^{-3}$,
$v_b/v_e= 6.0$, $g= 5.0\times 10^{-3}$, $\kappa_e= 5.0$.
}
\label{fig4}
\end{figure*}

In what follows, we investigate the time evolution of the beam-plasma
instability, comparing the situation in which the background electron
distribution is purely Maxwellian with a case in which there is also an
``halo'' population described by a Kappa distribution.
We
assume a beam population described by a displaced Maxwellian distribution,
with normalized beam  velocity $u_b=6.0$, number density given by
$n_{b}/n_e= 1.0\times 10^{-3}$, and temperature $T_b=T_e$,
\autoref{fig4} shows 2D plots of the electron velocity distribution.
Due to the presence of the beam, the background electron distribution
is slightly displaced in velocity space, so that the average velocity
of the complete electron velocity distribution is zero.

\autoref{fig4} shows 2D plots of the electron velocity distribution, the
spectrum of $L$ waves, and the spectrum of $T$ waves, at $\tau=500$. The
spectrum of $S$ waves remain very similar to the initial shape, and is
not shown. The three panels at the left column were obtained considering
that the background electron distribution is a Maxwellian distribution,
i.e., considering $n_{\kappa e}/n_e=0.0$, and the panels at the right 
column were obtained assuming $n_{\kappa e}/n_e= 0.05$, with
 $\kappa_e= 5.0$. For the parameters chosen, at such a point in the time
evolution the quasilinear process has already transferred to waves a
significant part of the energy available in the beam, creating a peak in
the spectrum of $L$ waves. The nonlinear processes are already operative,
creating a ring-like structure in the spectrum of $L$ waves, creating a
spectrum of $T$ waves over the whole grid of $q$ values, and creating
some peaked features for the $T$ waves, in the region of small values
of $q$. The situations depicted in Figures \ref{fig4}(a), (c) and (e)
correspond to those appearing in Figures 1(b), 2(b) and 4(b) of Ref.
\cite{pl:ZiebellYPGP15}, which is dedicated to the study of emission by
nonlinear processes in a plasma with Maxwellian background distributions.
It can be noticed in Figures \ref{fig4}(a) and \ref{fig4}(b) that the
region between the core of the velocity distribution and the peak of the
beam distribution is already quite flattened, corresponding to the
formation of the peak in the $L$ spectrum which is centered at
$(q_\perp,q_z)\simeq (0,0.2)$ in Figures \ref{fig4}(c) and \ref{fig4}(d).

\begin{figure*}
\includegraphics[width=0.9\textwidth]{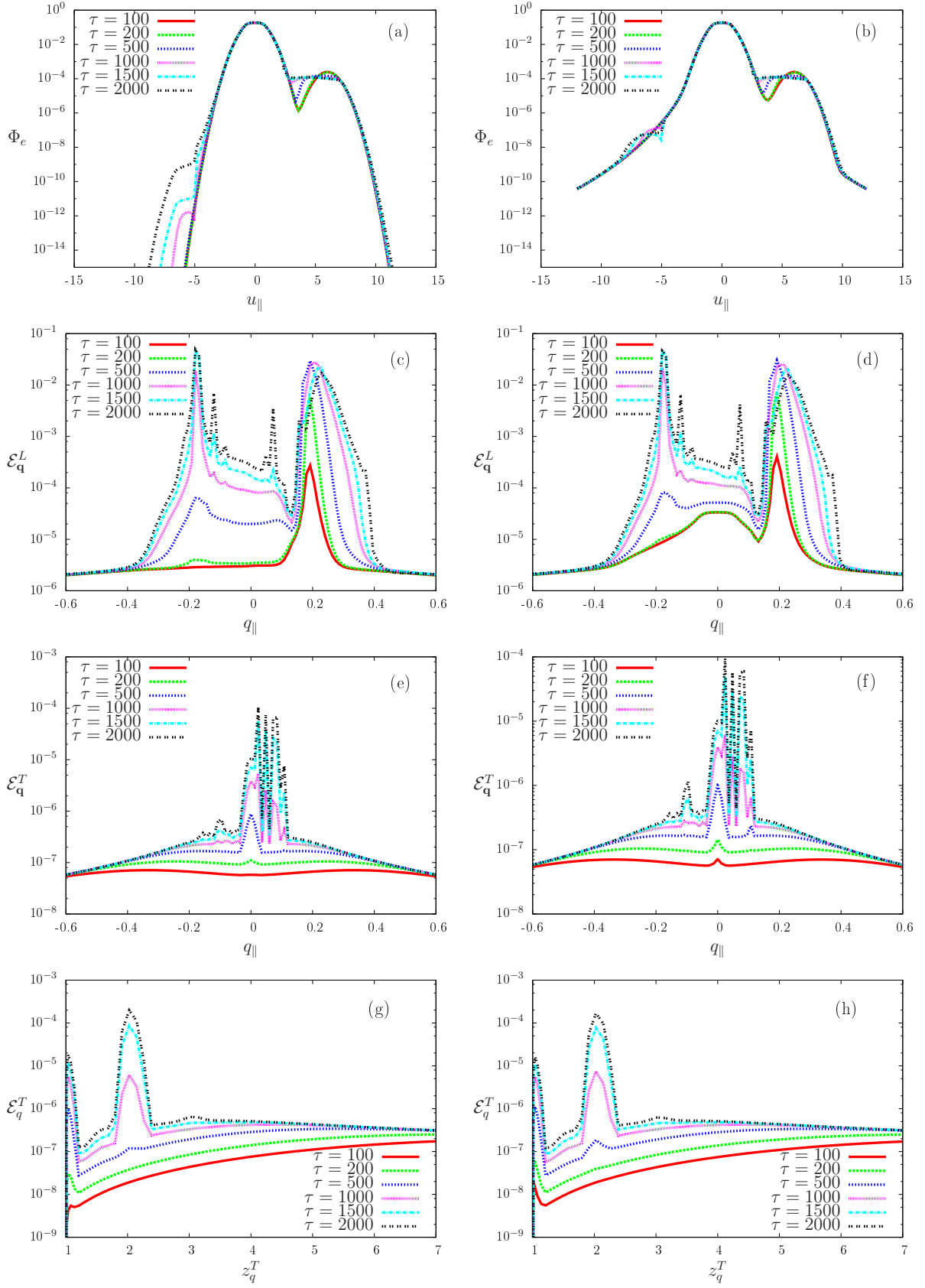}
\caption{
(a) 1D electron distribution function vs. $u_\parallel$, for several values
of $\tau$, for $n_{\kappa e}/n_e= 0.0$;
(b) 1D electron distribution function vs. $u_\parallel$. for several values
of $\tau=500$, for $n_{\kappa e}/n_e= 5.0\times 10^{-2}$;
(c) 1D spectrum of $L$ waves vs. $q_\parallel$, for several values of
$\tau$, for $n_{\kappa e}/n_e= 0.0$;
(d) 1D spectrum of $L$ waves vs. $q_\parallel$, for several values of $\tau$,
for $n_{\kappa e}/n_e= 5.0\times 10^{-2}$;
(e) 1D spectrum of $T$ waves vs. $q_\parallel$, for several values of $\tau$,
for $n_{\kappa e}/n_e= 0.0$;
(f) 1D spectrum of $T$ waves vs. $q_\parallel$, for several values of $\tau$,
for $n_{\kappa e}/n_e= 5.0\times 10^{-2}$.
(g) Spectrum of $T$ waves vs. $q$, for several values of $\tau$, for
$n_{\kappa e}/n_e= 0.0$;
(h) Spectrum of $T$ waves vs. $q$, for several values of $\tau$, for
$n_{\kappa e}/n_e= 5.0\times 10^{-2}$.
The parameters are the same used in \protect{\autoref{fig4}}.
}
\label{fig5}
\end{figure*}

The results appearing in \autoref{fig4} can be considered as
representative of the time evolution of the wave-particle system. For
further analysis of the time evolution, we show in \autoref{fig5}
1D representations of the electron distribution and of the wave spectra,
obtained after integration of the 2D quantities, along $u_\perp$ in the
case of the velocity distribution and along $q_\perp$ in the case of the
wave spectra. As in \autoref{fig4}, the left column displays results
obtained assuming $n_{\kappa e}/n_e= 0.0$, and the right column shows
results obtained assuming $n_{\kappa e}/n_e= 0.05$, with $\kappa_e=5.0$.
The electron distribution function in each case is shown in Figures
\ref{fig5}(a) and \ref{fig5}(b), respectively, for several values
of $\tau$, between $\tau=100$ and $\tau=2000$. In both panels, it can be
noticed the gradual flattening of the peak of the beam distribution, and the
formation of a \textit{plateau} in the region of velocities between the beam
and the core distribution. In \autoref{fig5}(a), it is also noticed the
appearing of a small population of backscattered electrons, which start to
become distinguishable at $\tau\simeq 1000$. In panel \ref{fig5}(b), these
backscattered electrons are not noticeable in the scale of the figure, because
the Kappa distribution already had a sizeable population at that region of
velocity space.

Figures \ref{fig5}(c) and \ref{fig5}(d) show a 1D projection of the $L$ wave
spectrum. In panel (c), one notices that at $\tau=100$ the only distinctive
feature in the spectrum is the primary peak generated at $q_z\simeq 0.2$, at
the spectral region where the waves are in resonance with electrons in the
region of positive velocity in the velocity distribution. At $\tau=200$,
there is already a hint of a backward peak, at $q_z\simeq -0.2$. At
$\tau= 500$, and beyond that, the backward peak appears well developed, and
there is a profile in the wave spectrum, continuous between the forward
peak and the backward peak. This is only a 1D projection of the ring formed
by scattering and decay, which is seen in the 2D representation appearing in
\autoref{fig4}(c). On the other hand, when the electron distribution
function features the presence of a Kappa distribution, the $L$ wave spectrum
at $\tau=100$ features the peak generated by quasilinear effect at $q_z\simeq
0.2$, and also the peak around $q=0$, characteristic of the spectrum at
equilibrium in the presence of a Kappa distribution. Due to the approximation
which we have adopted, of a limiting resonant velocity, the spectrum at
$q=0$ is finite instead of divergent. The 1D projection at \autoref{fig5}(d)
shows at $\tau=200$ already a hint of the backward peak. At $\tau=500$, and
beyond, the 1D spectrum of \autoref{fig5}(d) becomes
similar to that appearing at \autoref{fig5}(c), but this is only the effect
of the 1D projection. The actual spectrum in the case of \autoref{fig5}(d)
is constituted by the primary and the back-scattered peaks, by the peak around
$q=0$, and by the ring structure formed by nonlinear effects, as seen in
\autoref{fig4}(d).

The 1D projection of the spectrum of $T$ waves appears depicted in Figures
\ref{fig5}(e) and \ref{fig5}(f), for several values of $\tau$. In both
panels the sequence of lines show initially the formation of a background
spectrum of $T$ waves, added of the growth of a wave peak around $q_z=0$.
Between $\tau=500$ and $\tau=1000$, other peaked structures appear in the 1D
representations of Figures \ref{fig5}(e,f), which are projections of the
narrow ring structure seen in Figures \ref{fig4}(e) and \ref{fig4}(f). The
1D representations in \autoref{fig5}, as well as the 2D representations
in \autoref{fig4}, show that the $T$ wave spectrum obtained in the case
of Maxwellian electron distribution is very similar to the $T$ wave spectrum
obtained in the case of the presence of a ``halo'' described by a Kappa
distribution. The only noticeable difference is that the peaks appearing in
the $T$ wave spectrum are slightly higher in the case of $n_{\kappa e}\ne 0$,
panel (e), than in the case of $n_{\kappa e}=0$, panel (f),
for the same value of $\tau$.

Another representation of the $T$ wave spectrum appears in Figures
\ref{fig5}(g) and \ref{fig5}(h), which display the spectrum of $T$
waves after integration along pitch angle. That is, Figures
\ref{fig5}(g) and \ref{fig5}(h) show the quantity
\begin{displaymath}
{\cal E}_q= \int_0^{2\pi}d\theta\;q{\cal E}_{\bf q}^T
\end{displaymath}
as a function of the normalized wave frequency. This representation clearly
shows the early formation of the $T$ wave background, then the onset of the
primary peak of fundamental emission, with frequency equal to the electron
plasma frequency, and later on the onset of harmonic emission, with the
peak of emission at $2\omega_{pe}$ clearly emerging between $\tau=500$ and
$\tau=1000$. The comparison between \autoref{fig5}(h) and \autoref{fig5}(g)
show that the curves obtained in both cases are qualitatively the same, with
the sole difference that the peaks are slightly higher in the case of
$n_{\kappa e}\ne 0$, shown in \autoref{fig5}(h).

\section{Final remarks}
\label{sec:final}

In the present paper we have discussed the spectra of electrostatic and
electromagnetic waves which may be present at quiescent situation in plasmas
whose particles have velocity distribution functions which are a combination
of a thermal background and an energetic ``halo'' distribution.
The motivation for the study has been the abundance of measurements
made in the solar wind environment, by satellites at different orbits, which
show the occurrence of particle distribution functions with these
characteristics. For the analysis presented in the paper,
the electron velocity distribution has been represented as a summation of
a Maxwellian distribution function and an isotropic Kappa distribution, with
the fraction of population having the Kappa populations assumed as a free
parameter.

The investigation has been conducted using the theoretical framework of weak
turbulence theory. We have briefly discussed basic features of the equations
of weak turbulence theory, and we have initially used these equations to obtain
expressions for the spectra of electrostatic waves,
obtained as the outcome of the
balance between spontaneous fluctuations and induced emission. These
equilibrium spectra, for high frequency Langmuir waves ($L$) and for low
frequency ion-acoustic waves ($S$), have been routinely discussed in the
literature for the case of
Maxwellian plasmas, but the present paper presents as a novel feature a
description of the effects of the presence of a population of particles
described by a Kappa velocity distribution. Theoretical expressions for the
spectra of L and S waves have been obtained considering that both ions and
electrons can be described by a combination of Maxwellian and Kappa
distribution. Some numerical results have also been presented, considering
the case of Maxwellian distribution for the ions and the combined distribution
for electrons, and considering different values of the $\kappa_e$ index.
These results show that the effect of the presence of the Kappa
distribution is noticeable in the spectrum of $L$ waves in the region of large
wavelengths, with difference relative to the spectrum obtained in the case
of purely Maxwellian distribution which increases for decreasing values of
the index $\kappa_e$ in the energetic population. The distinctive feature,
which exists even for very tenuous Kappa population, is the presence of a peak
of wave intensity for very large wavelengths (wave-number $k\to 0$).

We have also discussed the characteristics of the spectrum of electromagnetic
waves ($T$), which shall be present in the plasma as the outcome of nonlinear
processes involving $L$ and $S$ waves, and particles. These spectra can be
characterized as a state of ``turbulent equilibrium''. The turbulent
equilibrium spectra have already been discussed for the case of Maxwellian
velocity distributions, and the present paper extends the discussion for
the case in which an energetic ``halo'' described as a Kappa distribution is
also present in the plasma. The results obtained show that the spectrum of
$T$ waves has the general features similar to those obtained in the case of
Maxwellian distributions, with the effect of the presence of the Kappa
population appearing as a peak of $T$ waves in the large wavelength region,
much narrower than the peak obtained in the spectrum of the $L$ waves.

In addition to the results concerning the equilibrium spectra, we have
also presented some results which show the time evolution of the spectra of
$L$ and $T$ waves, and the time evolution of the electron distribution
function, as a result of the presence of a tenuous electron beam travelling in
the plasma. We have followed the time evolution of the wave-particle system
up to the formation of the \textit{plateau} in the electron distribution
function which indicates the saturation of the induced processes described
by quasilinear theory. The results which were shown in the paper compare the
results obtained in the case in which the background electron population is
described by a Maxwellian distribution, with results obtained in the case of a
background distribution described as a core population with Maxwellian
distribution and a tenuous population with isotropic Kappa distribution.
It was shown that the time evolution of the spectrum of $L$ waves
obtained in the presence of the ``halo'' distribution is qualitatively very
similar to the spectrum obtained in the case of thermal background
distribution, except for the occurrence of the enhanced wave intensity for
$k\to 0$, characteristic of the presence of a Kappa population of particles.
The spectra obtained for the $T$ waves along the time evolution, in the
two situations which have been considered, are also qualitatively very similar,
with the difference that the peak corresponding to the harmonic emission
is slightly more pronounced in the presence of a tenuous Kappa distribution,
in comparison with harmonic emission obtained in the case of Maxwellian
background population.


\clearpage{}

\begin{acknowledgments}
SFT and LTP acknowledge PhD fellowships from CNPq (Brazil).
LFZ acknowledges support from CNPq (Brazil), grant No. 304363/2014-6.
RG acknowledges support from CNPq (Brazil), grants No. 478728/2012-3
and 307626/2015-6.
PHY acknowledges NSF grant AGS1550566 to the University
of Maryland, the BK21 plus program from the National
Research Foundation (NRF), Korea, to Kyung Hee University,
and the Science Award Grant from the GFT, Inc., to the
University of Maryland.
\end{acknowledgments}


\bibliographystyle{aipnum4-1}
\bibliography{references}

\providecommand{\noopsort}[1]{}\providecommand{\singleletter}[1]{#1}%
\begin{thebibliography}{37}%
\makeatletter
\providecommand \@ifxundefined [1]{%
 \@ifx{#1\undefined}
}%
\providecommand \@ifnum [1]{%
 \ifnum #1\expandafter \@firstoftwo
 \else \expandafter \@secondoftwo
 \fi
}%
\providecommand \@ifx [1]{%
 \ifx #1\expandafter \@firstoftwo
 \else \expandafter \@secondoftwo
 \fi
}%
\providecommand \natexlab [1]{#1}%
\providecommand \enquote  [1]{``#1''}%
\providecommand \bibnamefont  [1]{#1}%
\providecommand \bibfnamefont [1]{#1}%
\providecommand \citenamefont [1]{#1}%
\providecommand \href@noop [0]{\@secondoftwo}%
\providecommand \href [0]{\begingroup \@sanitize@url \@href}%
\providecommand \@href[1]{\@@startlink{#1}\@@href}%
\providecommand \@@href[1]{\endgroup#1\@@endlink}%
\providecommand \@sanitize@url [0]{\catcode `\\12\catcode `\$12\catcode
  `\&12\catcode `\#12\catcode `\^12\catcode `\_12\catcode `\%12\relax}%
\providecommand \@@startlink[1]{}%
\providecommand \@@endlink[0]{}%
\providecommand \url  [0]{\begingroup\@sanitize@url \@url }%
\providecommand \@url [1]{\endgroup\@href {#1}{\urlprefix }}%
\providecommand \urlprefix  [0]{URL }%
\providecommand \Eprint [0]{\href }%
\providecommand \doibase [0]{http://dx.doi.org/}%
\providecommand \selectlanguage [0]{\@gobble}%
\providecommand \bibinfo  [0]{\@secondoftwo}%
\providecommand \bibfield  [0]{\@secondoftwo}%
\providecommand \translation [1]{[#1]}%
\providecommand \BibitemOpen [0]{}%
\providecommand \bibitemStop [0]{}%
\providecommand \bibitemNoStop [0]{.\EOS\space}%
\providecommand \EOS [0]{\spacefactor3000\relax}%
\providecommand \BibitemShut  [1]{\csname bibitem#1\endcsname}%
\let\auto@bib@innerbib\@empty
\bibitem [{\citenamefont {Feldman}\ \emph {et~al.}(1975)\citenamefont
  {Feldman}, \citenamefont {Asbridge}, \citenamefont {Bame}, \citenamefont
  {Montgomery},\ and\ \citenamefont {Gary}}]{Feldman+75}%
  \BibitemOpen
  \bibfield  {author} {\bibinfo {author} {\bibfnamefont {W.~C.}\ \bibnamefont
  {Feldman}}, \bibinfo {author} {\bibfnamefont {J.~R.}\ \bibnamefont
  {Asbridge}}, \bibinfo {author} {\bibfnamefont {S.~J.}\ \bibnamefont {Bame}},
  \bibinfo {author} {\bibfnamefont {M.~D.}\ \bibnamefont {Montgomery}}, \ and\
  \bibinfo {author} {\bibfnamefont {S.~P.}\ \bibnamefont {Gary}},\ }\href
  {\doibase 10.1029/JA080i031p04181} {\bibfield  {journal} {\bibinfo  {journal}
  {J. Geophys. Res.}\ }\textbf {\bibinfo {volume} {80}},\ \bibinfo {pages}
  {4181} (\bibinfo {year} {1975})}\BibitemShut {NoStop}%
\bibitem [{\citenamefont {Pilipp}\ \emph {et~al.}(1987)\citenamefont {Pilipp},
  \citenamefont {Miggenrieder}, \citenamefont {Montgomery}, \citenamefont
  {M\"uhlh\"auser}, \citenamefont {Rosenbauer},\ and\ \citenamefont
  {Schwenn}}]{Pilipp+87}%
  \BibitemOpen
  \bibfield  {author} {\bibinfo {author} {\bibfnamefont {W.~G.}\ \bibnamefont
  {Pilipp}}, \bibinfo {author} {\bibfnamefont {H.}~\bibnamefont
  {Miggenrieder}}, \bibinfo {author} {\bibfnamefont {M.~D.}\ \bibnamefont
  {Montgomery}}, \bibinfo {author} {\bibfnamefont {K.-H.}\ \bibnamefont
  {M\"uhlh\"auser}}, \bibinfo {author} {\bibfnamefont {H.}~\bibnamefont
  {Rosenbauer}}, \ and\ \bibinfo {author} {\bibfnamefont {R.}~\bibnamefont
  {Schwenn}},\ }\href {\doibase 10.1029/JA092iA02p01075} {\bibfield  {journal}
  {\bibinfo  {journal} {J. Geophys. Res.}\ }\textbf {\bibinfo {volume} {92}},\
  \bibinfo {pages} {1075} (\bibinfo {year} {1987})}\BibitemShut {NoStop}%
\bibitem [{\citenamefont {Lin}\ \emph {et~al.}(1995)\citenamefont {Lin},
  \citenamefont {Anderson}, \citenamefont {Ashford}, \citenamefont {Carlson},
  \citenamefont {Curtis}, \citenamefont {Ergun}, \citenamefont {Larson},
  \citenamefont {McFadden}, \citenamefont {McCarthy}, \citenamefont {Parks},
  \citenamefont {R\'eme}, \citenamefont {Bosqued}, \citenamefont {Coutelier},
  \citenamefont {Cotin}, \citenamefont {D'Uston}, \citenamefont {Wenzel},
  \citenamefont {Sanderson}, \citenamefont {Henrion}, \citenamefont {Ronnet},\
  and\ \citenamefont {Paschmann}}]{Lin+95}%
  \BibitemOpen
  \bibfield  {author} {\bibinfo {author} {\bibfnamefont {R.~P.}\ \bibnamefont
  {Lin}}, \bibinfo {author} {\bibfnamefont {K.~A.}\ \bibnamefont {Anderson}},
  \bibinfo {author} {\bibfnamefont {S.}~\bibnamefont {Ashford}}, \bibinfo
  {author} {\bibfnamefont {C.}~\bibnamefont {Carlson}}, \bibinfo {author}
  {\bibfnamefont {D.}~\bibnamefont {Curtis}}, \bibinfo {author} {\bibfnamefont
  {R.}~\bibnamefont {Ergun}}, \bibinfo {author} {\bibfnamefont
  {D.}~\bibnamefont {Larson}}, \bibinfo {author} {\bibfnamefont
  {J.}~\bibnamefont {McFadden}}, \bibinfo {author} {\bibfnamefont
  {M.}~\bibnamefont {McCarthy}}, \bibinfo {author} {\bibfnamefont {G.~K.}\
  \bibnamefont {Parks}}, \bibinfo {author} {\bibfnamefont {H.}~\bibnamefont
  {R\'eme}}, \bibinfo {author} {\bibfnamefont {J.~M.}\ \bibnamefont {Bosqued}},
  \bibinfo {author} {\bibfnamefont {J.}~\bibnamefont {Coutelier}}, \bibinfo
  {author} {\bibfnamefont {F.}~\bibnamefont {Cotin}}, \bibinfo {author}
  {\bibfnamefont {C.}~\bibnamefont {D'Uston}}, \bibinfo {author} {\bibfnamefont
  {K.-P.}\ \bibnamefont {Wenzel}}, \bibinfo {author} {\bibfnamefont {T.~R.}\
  \bibnamefont {Sanderson}}, \bibinfo {author} {\bibfnamefont {J.}~\bibnamefont
  {Henrion}}, \bibinfo {author} {\bibfnamefont {J.~C.}\ \bibnamefont {Ronnet}},
  \ and\ \bibinfo {author} {\bibfnamefont {G.}~\bibnamefont {Paschmann}},\
  }\href {\doibase 10.1007/BF00751328} {\bibfield  {journal} {\bibinfo
  {journal} {Space Sci. Rev.}\ }\textbf {\bibinfo {volume} {71}},\ \bibinfo
  {pages} {125} (\bibinfo {year} {1995})}\BibitemShut {NoStop}%
\bibitem [{\citenamefont {Stone}, \citenamefont {Summings},\ and\ \citenamefont
  {McDonald}(2008)}]{Stone+08}%
  \BibitemOpen
  \bibfield  {author} {\bibinfo {author} {\bibfnamefont {E.~C.}\ \bibnamefont
  {Stone}}, \bibinfo {author} {\bibfnamefont {A.~C.}\ \bibnamefont {Summings}},
  \ and\ \bibinfo {author} {\bibfnamefont {F.~B.}\ \bibnamefont {McDonald}},\
  }\href {\doibase 10.1007/BF00751328} {\bibfield  {journal} {\bibinfo
  {journal} {Nature}\ }\textbf {\bibinfo {volume} {454}},\ \bibinfo {pages}
  {71} (\bibinfo {year} {2008})}\BibitemShut {NoStop}%
\bibitem [{\citenamefont {Krucker}\ and\ \citenamefont
  {Battaglia}(2014)}]{KruckerBattaglia14}%
  \BibitemOpen
  \bibfield  {author} {\bibinfo {author} {\bibfnamefont {S.}~\bibnamefont
  {Krucker}}\ and\ \bibinfo {author} {\bibfnamefont {M.}~\bibnamefont
  {Battaglia}},\ }\href {\doibase 10.1088/0004-637X/780/1/107} {\bibfield
  {journal} {\bibinfo  {journal} {Astrophys. J.}\ }\textbf {\bibinfo {volume}
  {780}},\ \bibinfo {pages} {107} (\bibinfo {year} {2014})}\BibitemShut
  {NoStop}%
\bibitem [{\citenamefont {Oka}\ \emph {et~al.}(2015)\citenamefont {Oka},
  \citenamefont {Krucker}, \citenamefont {Hudson},\ and\ \citenamefont
  {Saint-Hilaire}}]{Oka+15}%
  \BibitemOpen
  \bibfield  {author} {\bibinfo {author} {\bibfnamefont {M.}~\bibnamefont
  {Oka}}, \bibinfo {author} {\bibfnamefont {S.}~\bibnamefont {Krucker}},
  \bibinfo {author} {\bibfnamefont {H.~S.}\ \bibnamefont {Hudson}}, \ and\
  \bibinfo {author} {\bibfnamefont {P.}~\bibnamefont {Saint-Hilaire}},\ }\href
  {\doibase 10.1088/0004-637X/799/2/129} {\bibfield  {journal} {\bibinfo
  {journal} {Astrophys. J.}\ }\textbf {\bibinfo {volume} {799}},\ \bibinfo
  {pages} {129} (\bibinfo {year} {2015})}\BibitemShut {NoStop}%
\bibitem [{\citenamefont {Wang}\ \emph {et~al.}(2012)\citenamefont {Wang},
  \citenamefont {Lin}, \citenamefont {Salem}, \citenamefont {Pulupa},
  \citenamefont {Larson}, \citenamefont {Yoon},\ and\ \citenamefont
  {Luhmann}}]{Wang+12}%
  \BibitemOpen
  \bibfield  {author} {\bibinfo {author} {\bibfnamefont {L.}~\bibnamefont
  {Wang}}, \bibinfo {author} {\bibfnamefont {R.~P.}\ \bibnamefont {Lin}},
  \bibinfo {author} {\bibfnamefont {C.}~\bibnamefont {Salem}}, \bibinfo
  {author} {\bibfnamefont {M.}~\bibnamefont {Pulupa}}, \bibinfo {author}
  {\bibfnamefont {D.~E.}\ \bibnamefont {Larson}}, \bibinfo {author}
  {\bibfnamefont {P.~H.}\ \bibnamefont {Yoon}}, \ and\ \bibinfo {author}
  {\bibfnamefont {J.~G.}\ \bibnamefont {Luhmann}},\ }\href {\doibase
  10.1088/2041-8205/753/1/L23} {\bibfield  {journal} {\bibinfo  {journal}
  {Astrophys. J. Lett.}\ }\textbf {\bibinfo {volume} {753}},\ \bibinfo {pages}
  {L23} (\bibinfo {year} {2012})}\BibitemShut {NoStop}%
\bibitem [{\citenamefont {Vasyliunas}(1968)}]{Vasyliunas68}%
  \BibitemOpen
  \bibfield  {author} {\bibinfo {author} {\bibfnamefont {V.~M.}\ \bibnamefont
  {Vasyliunas}},\ }\href {\doibase 10.1029/JA073i009p02839} {\bibfield
  {journal} {\bibinfo  {journal} {J. Geophys. Res.}\ }\textbf {\bibinfo
  {volume} {73}},\ \bibinfo {pages} {2839} (\bibinfo {year}
  {1968})}\BibitemShut {NoStop}%
\bibitem [{\citenamefont {Summers}\ and\ \citenamefont
  {Thorne}(1991)}]{SummersT91}%
  \BibitemOpen
  \bibfield  {author} {\bibinfo {author} {\bibfnamefont {D.}~\bibnamefont
  {Summers}}\ and\ \bibinfo {author} {\bibfnamefont {R.~M.}\ \bibnamefont
  {Thorne}},\ }\href {\doibase 10.1063/1.859653} {\bibfield  {journal}
  {\bibinfo  {journal} {Phys. Fluids B}\ }\textbf {\bibinfo {volume} {3}},\
  \bibinfo {pages} {1835} (\bibinfo {year} {1991})}\BibitemShut {NoStop}%
\bibitem [{\citenamefont {Mace}\ and\ \citenamefont
  {Hellberg}(1995)}]{MaceHellberg95}%
  \BibitemOpen
  \bibfield  {author} {\bibinfo {author} {\bibfnamefont {R.~L.}\ \bibnamefont
  {Mace}}\ and\ \bibinfo {author} {\bibfnamefont {M.~A.}\ \bibnamefont
  {Hellberg}},\ }\href {\doibase 10.1063/1.871296} {\bibfield  {journal}
  {\bibinfo  {journal} {Phys. Plasmas}\ }\textbf {\bibinfo {volume} {2}},\
  \bibinfo {pages} {2098} (\bibinfo {year} {1995})}\BibitemShut {NoStop}%
\bibitem [{\citenamefont {Leubner}\ and\ \citenamefont
  {Schupfer}(2000)}]{LeubnerS00}%
  \BibitemOpen
  \bibfield  {author} {\bibinfo {author} {\bibfnamefont {M.~P.}\ \bibnamefont
  {Leubner}}\ and\ \bibinfo {author} {\bibfnamefont {N.}~\bibnamefont
  {Schupfer}},\ }\href {\doibase 10.1029/1999JA000447} {\bibfield  {journal}
  {\bibinfo  {journal} {J. Geophys. Res.}\ }\textbf {\bibinfo {volume} {105}},\
  \bibinfo {pages} {27387} (\bibinfo {year} {2000})}\BibitemShut {NoStop}%
\bibitem [{\citenamefont {Leubner}\ and\ \citenamefont
  {Schupfer}(2001)}]{LeubnerS01}%
  \BibitemOpen
  \bibfield  {author} {\bibinfo {author} {\bibfnamefont {M.~P.}\ \bibnamefont
  {Leubner}}\ and\ \bibinfo {author} {\bibfnamefont {N.}~\bibnamefont
  {Schupfer}},\ }\href {\doibase 10.1029/2000JA000425} {\bibfield  {journal}
  {\bibinfo  {journal} {J. Geophys. Res.}\ }\textbf {\bibinfo {volume} {106}},\
  \bibinfo {pages} {12993} (\bibinfo {year} {2001})}\BibitemShut {NoStop}%
\bibitem [{\citenamefont {Leubner}(2002)}]{Leubner02}%
  \BibitemOpen
  \bibfield  {author} {\bibinfo {author} {\bibfnamefont {M.~P.}\ \bibnamefont
  {Leubner}},\ }\href {\doibase 10.1023/A:1020990413487} {\bibfield  {journal}
  {\bibinfo  {journal} {Astrophys. Space Sci.}\ }\textbf {\bibinfo {volume}
  {282}},\ \bibinfo {pages} {573} (\bibinfo {year} {2002})}\BibitemShut
  {NoStop}%
\bibitem [{\citenamefont {Leubner}(2004)}]{Leubner04b}%
  \BibitemOpen
  \bibfield  {author} {\bibinfo {author} {\bibfnamefont {M.~P.}\ \bibnamefont
  {Leubner}},\ }\href {\doibase 10.1086/381867} {\bibfield  {journal} {\bibinfo
   {journal} {Astrophys. J.}\ }\textbf {\bibinfo {volume} {604}},\ \bibinfo
  {pages} {469} (\bibinfo {year} {2004})}\BibitemShut {NoStop}%
\bibitem [{\citenamefont {Kim}\ \emph {et~al.}(2015)\citenamefont {Kim},
  \citenamefont {Yoon}, \citenamefont {Choe},\ and\ \citenamefont
  {Wang}}]{Kim+15}%
  \BibitemOpen
  \bibfield  {author} {\bibinfo {author} {\bibfnamefont {S.}~\bibnamefont
  {Kim}}, \bibinfo {author} {\bibfnamefont {P.~H.}\ \bibnamefont {Yoon}},
  \bibinfo {author} {\bibfnamefont {G.~S.}\ \bibnamefont {Choe}}, \ and\
  \bibinfo {author} {\bibfnamefont {L.}~\bibnamefont {Wang}},\ }\href {\doibase
  10.1088/0004-637X/806/1/32} {\bibfield  {journal} {\bibinfo  {journal}
  {Astrophys. J.}\ }\textbf {\bibinfo {volume} {806}},\ \bibinfo {pages} {32}
  (\bibinfo {year} {2015})}\BibitemShut {NoStop}%
\bibitem [{\citenamefont {McComas}\ \emph {et~al.}(2003)\citenamefont
  {McComas}, \citenamefont {Elliott}, \citenamefont {Schwadron}, \citenamefont
  {Gosling}, \citenamefont {Skoug},\ and\ \citenamefont
  {Goldstein}}]{McComas+03}%
  \BibitemOpen
  \bibfield  {author} {\bibinfo {author} {\bibfnamefont {D.~J.}\ \bibnamefont
  {McComas}}, \bibinfo {author} {\bibfnamefont {H.~A.}\ \bibnamefont
  {Elliott}}, \bibinfo {author} {\bibfnamefont {N.~A.}\ \bibnamefont
  {Schwadron}}, \bibinfo {author} {\bibfnamefont {J.~T.}\ \bibnamefont
  {Gosling}}, \bibinfo {author} {\bibfnamefont {R.~M.}\ \bibnamefont {Skoug}},
  \ and\ \bibinfo {author} {\bibfnamefont {B.~E.}\ \bibnamefont {Goldstein}},\
  }\href {\doibase 10.1029/2003GL017136} {\bibfield  {journal} {\bibinfo
  {journal} {Geophys. Res. Lett.}\ }\textbf {\bibinfo {volume} {30}},\ \bibinfo
  {pages} {24} (\bibinfo {year} {2003})}\BibitemShut {NoStop}%
\bibitem [{\citenamefont {Maksimovic}\ \emph {et~al.}(2005)\citenamefont
  {Maksimovic}, \citenamefont {Zouganelis}, \citenamefont {Chaufrey},
  \citenamefont {Issautier}, \citenamefont {Scime}, \citenamefont {Littleton},
  \citenamefont {Marsch}, \citenamefont {McComas}, \citenamefont {Salem},
  \citenamefont {Lin},\ and\ \citenamefont {Elliott}}]{Maksimovic+05}%
  \BibitemOpen
  \bibfield  {author} {\bibinfo {author} {\bibfnamefont {M.}~\bibnamefont
  {Maksimovic}}, \bibinfo {author} {\bibfnamefont {I.}~\bibnamefont
  {Zouganelis}}, \bibinfo {author} {\bibfnamefont {J.-Y.}\ \bibnamefont
  {Chaufrey}}, \bibinfo {author} {\bibfnamefont {K.}~\bibnamefont {Issautier}},
  \bibinfo {author} {\bibfnamefont {E.~E.}\ \bibnamefont {Scime}}, \bibinfo
  {author} {\bibfnamefont {J.~E.}\ \bibnamefont {Littleton}}, \bibinfo {author}
  {\bibfnamefont {E.}~\bibnamefont {Marsch}}, \bibinfo {author} {\bibfnamefont
  {D.~J.}\ \bibnamefont {McComas}}, \bibinfo {author} {\bibfnamefont
  {C.}~\bibnamefont {Salem}}, \bibinfo {author} {\bibfnamefont {R.~P.}\
  \bibnamefont {Lin}}, \ and\ \bibinfo {author} {\bibfnamefont
  {H.}~\bibnamefont {Elliott}},\ }\href {\doibase 10.1029/2005JA011119}
  {\bibfield  {journal} {\bibinfo  {journal} {J. Geophys. Res.}\ }\textbf
  {\bibinfo {volume} {110}},\ \bibinfo {pages} {A09104} (\bibinfo {year}
  {2005})}\BibitemShut {NoStop}%
\bibitem [{\citenamefont {Vocks}\ and\ \citenamefont
  {Mann}(2003)}]{VocksMann03}%
  \BibitemOpen
  \bibfield  {author} {\bibinfo {author} {\bibfnamefont {C.}~\bibnamefont
  {Vocks}}\ and\ \bibinfo {author} {\bibfnamefont {G.}~\bibnamefont {Mann}},\
  }\href {\doibase 10.1086/376682} {\bibfield  {journal} {\bibinfo  {journal}
  {Astrophys. J.}\ }\textbf {\bibinfo {volume} {593}},\ \bibinfo {pages} {1134}
  (\bibinfo {year} {2003})}\BibitemShut {NoStop}%
\bibitem [{\citenamefont {G.~Livadiotis}(2017)}]{LivadiotisBook}%
  \BibitemOpen
  \bibfield  {author} {\bibinfo {author} {\bibfnamefont {E.}~\bibnamefont
  {G.~Livadiotis}},\ }\href@noop {} {\emph {\bibinfo {title} {Kappa
  Distributions}}}\ (\bibinfo  {publisher} {Elsevier},\ \bibinfo {address}
  {Amsterdam},\ \bibinfo {year} {2017})\BibitemShut {NoStop}%
\bibitem [{\citenamefont {Hellberg}\ \emph {et~al.}(2009)\citenamefont
  {Hellberg}, \citenamefont {Mace}, \citenamefont {Baluku}, \citenamefont
  {Kourakis},\ and\ \citenamefont {Saini}}]{HellbergMBKS09}%
  \BibitemOpen
  \bibfield  {author} {\bibinfo {author} {\bibfnamefont {M.~A.}\ \bibnamefont
  {Hellberg}}, \bibinfo {author} {\bibfnamefont {R.~L.}\ \bibnamefont {Mace}},
  \bibinfo {author} {\bibfnamefont {T.~K.}\ \bibnamefont {Baluku}}, \bibinfo
  {author} {\bibfnamefont {I.}~\bibnamefont {Kourakis}}, \ and\ \bibinfo
  {author} {\bibfnamefont {N.~S.}\ \bibnamefont {Saini}},\ }\href {\doibase
  10.1063/1.3213388} {\bibfield  {journal} {\bibinfo  {journal} {Phys.
  Plasmas}\ }\textbf {\bibinfo {volume} {16}},\ \bibinfo {pages} {094701}
  (\bibinfo {year} {2009})}\BibitemShut {NoStop}%
\bibitem [{\citenamefont {Hapgood}\ \emph {et~al.}(2011)\citenamefont
  {Hapgood}, \citenamefont {Perry}, \citenamefont {Davies},\ and\ \citenamefont
  {Denton}}]{HapgoodPDD11}%
  \BibitemOpen
  \bibfield  {author} {\bibinfo {author} {\bibfnamefont {M.}~\bibnamefont
  {Hapgood}}, \bibinfo {author} {\bibfnamefont {C.}~\bibnamefont {Perry}},
  \bibinfo {author} {\bibfnamefont {J.}~\bibnamefont {Davies}}, \ and\ \bibinfo
  {author} {\bibfnamefont {M.}~\bibnamefont {Denton}},\ }\href {\doibase
  10.1016/j.pss.2010.06.002} {\bibfield  {journal} {\bibinfo  {journal}
  {Planet. Space Sci.}\ }\textbf {\bibinfo {volume} {59}},\ \bibinfo {pages}
  {618} (\bibinfo {year} {2011})}\BibitemShut {NoStop}%
\bibitem [{\citenamefont {Livadiotis}\ and\ \citenamefont
  {McComas}(2013)}]{LivadiotisM13}%
  \BibitemOpen
  \bibfield  {author} {\bibinfo {author} {\bibfnamefont {G.}~\bibnamefont
  {Livadiotis}}\ and\ \bibinfo {author} {\bibfnamefont {D.~J.}\ \bibnamefont
  {McComas}},\ }\href {\doibase 10.1007/s11214-013-9982-9} {\bibfield
  {journal} {\bibinfo  {journal} {Space Sci. Rev.}\ }\textbf {\bibinfo {volume}
  {175}},\ \bibinfo {pages} {183} (\bibinfo {year} {2013})}\BibitemShut
  {NoStop}%
\bibitem [{\citenamefont {Livadiotis}(2015)}]{Livadiotis2015}%
  \BibitemOpen
  \bibfield  {author} {\bibinfo {author} {\bibfnamefont {G.}~\bibnamefont
  {Livadiotis}},\ }\href {\doibase 10.1002/2014JA020825} {\bibfield  {journal}
  {\bibinfo  {journal} {J. Geophys. Res.}\ }\textbf {\bibinfo {volume} {120}},\
  \bibinfo {pages} {1607} (\bibinfo {year} {2015})}\BibitemShut {NoStop}%
\bibitem [{\citenamefont {Lazar}, \citenamefont {Fichtner},\ and\ \citenamefont
  {Yoon}(2016)}]{LazarFY16}%
  \BibitemOpen
  \bibfield  {author} {\bibinfo {author} {\bibfnamefont {M.}~\bibnamefont
  {Lazar}}, \bibinfo {author} {\bibfnamefont {H.}~\bibnamefont {Fichtner}}, \
  and\ \bibinfo {author} {\bibfnamefont {P.~H.}\ \bibnamefont {Yoon}},\ }\href
  {\doibase 10.1051/0004-6361/201527593} {\bibfield  {journal} {\bibinfo
  {journal} {Astron. Astrophys.}\ }\textbf {\bibinfo {volume} {589}},\ \bibinfo
  {pages} {A39} (\bibinfo {year} {2016})}\BibitemShut {NoStop}%
\bibitem [{\citenamefont {Yoon}(2012)}]{Yoon2012}%
  \BibitemOpen
  \bibfield  {author} {\bibinfo {author} {\bibfnamefont {P.~H.}\ \bibnamefont
  {Yoon}},\ }\href {\doibase 10.1063/1.4710515} {\bibfield  {journal} {\bibinfo
   {journal} {Physics of Plasmas}\ }\textbf {\bibinfo {volume} {19}},\ \bibinfo
  {pages} {052301} (\bibinfo {year} {2012})}\BibitemShut {NoStop}%
\bibitem [{\citenamefont {Yoon}(2014)}]{Yoon2014}%
  \BibitemOpen
  \bibfield  {author} {\bibinfo {author} {\bibfnamefont {P.~H.}\ \bibnamefont
  {Yoon}},\ }\href {\doibase 10.1002/2014JA020353} {\bibfield  {journal}
  {\bibinfo  {journal} {J. Geophys. Res.}\ }\textbf {\bibinfo {volume} {119}},\
  \bibinfo {pages} {7074} (\bibinfo {year} {2014})}\BibitemShut {NoStop}%
\bibitem [{\citenamefont {Kim}\ \emph {et~al.}(2016)\citenamefont {Kim},
  \citenamefont {Yoon}, \citenamefont {Choe},\ and\ \citenamefont
  {moon}}]{Kim+16}%
  \BibitemOpen
  \bibfield  {author} {\bibinfo {author} {\bibfnamefont {S.}~\bibnamefont
  {Kim}}, \bibinfo {author} {\bibfnamefont {P.~H.}\ \bibnamefont {Yoon}},
  \bibinfo {author} {\bibfnamefont {G.~S.}\ \bibnamefont {Choe}}, \ and\
  \bibinfo {author} {\bibfnamefont {Y.-J.}\ \bibnamefont {moon}},\ }\href
  {http://stacks.iop.org/0004-637X/828/i=1/a=60} {\bibfield  {journal}
  {\bibinfo  {journal} {Astrophys. J.}\ }\textbf {\bibinfo {volume} {828}},\
  \bibinfo {pages} {60} (\bibinfo {year} {2016})}\BibitemShut {NoStop}%
\bibitem [{\citenamefont {Yoon}\ \emph {et~al.}(2012)\citenamefont {Yoon},
  \citenamefont {Ziebell}, \citenamefont {Gaelzer},\ and\ \citenamefont
  {Pavan}}]{pl:YoonZGP12}%
  \BibitemOpen
  \bibfield  {author} {\bibinfo {author} {\bibfnamefont {P.~H.}\ \bibnamefont
  {Yoon}}, \bibinfo {author} {\bibfnamefont {L.~F.}\ \bibnamefont {Ziebell}},
  \bibinfo {author} {\bibfnamefont {R.}~\bibnamefont {Gaelzer}}, \ and\
  \bibinfo {author} {\bibfnamefont {J.}~\bibnamefont {Pavan}},\ }\href
  {\doibase 10.1063/1.4757224} {\bibfield  {journal} {\bibinfo  {journal}
  {Phys. Plasmas}\ }\textbf {\bibinfo {volume} {19}},\ \bibinfo {pages}
  {102303, 9pp} (\bibinfo {year} {2012})}\BibitemShut {NoStop}%
\bibitem [{\citenamefont {Tsallis}(1988)}]{Tsallis88}%
  \BibitemOpen
  \bibfield  {author} {\bibinfo {author} {\bibfnamefont {C.}~\bibnamefont
  {Tsallis}},\ }\href {\doibase 10.1007/BF01016429} {\bibfield  {journal}
  {\bibinfo  {journal} {J. Stat. Phys.}\ }\textbf {\bibinfo {volume} {52}},\
  \bibinfo {pages} {479} (\bibinfo {year} {1988})}\BibitemShut {NoStop}%
\bibitem [{\citenamefont {Silva}, \citenamefont {Plastino},\ and\ \citenamefont
  {Lima}(1998)}]{SilvaPL98}%
  \BibitemOpen
  \bibfield  {author} {\bibinfo {author} {\bibfnamefont {R.}~\bibnamefont
  {Silva}}, \bibinfo {author} {\bibfnamefont {A.~R.}\ \bibnamefont {Plastino}},
  \ and\ \bibinfo {author} {\bibfnamefont {J.~A.~S.}\ \bibnamefont {Lima}},\
  }\href {\doibase 10.1016/S0375-9601(98)00710-5} {\bibfield  {journal}
  {\bibinfo  {journal} {Phys. Lett. A}\ }\textbf {\bibinfo {volume} {249}},\
  \bibinfo {pages} {401} (\bibinfo {year} {1998})}\BibitemShut {NoStop}%
\bibitem [{\citenamefont {Tsallis}, \citenamefont {Mendes},\ and\ \citenamefont
  {Plastino}(1998)}]{TsallisMP98}%
  \BibitemOpen
  \bibfield  {author} {\bibinfo {author} {\bibfnamefont {C.}~\bibnamefont
  {Tsallis}}, \bibinfo {author} {\bibfnamefont {R.~S.}\ \bibnamefont {Mendes}},
  \ and\ \bibinfo {author} {\bibfnamefont {A.~R.}\ \bibnamefont {Plastino}},\
  }\href {\doibase 10.1016/S0378-4371(98)00437-3} {\bibfield  {journal}
  {\bibinfo  {journal} {Phys. A}\ }\textbf {\bibinfo {volume} {261}},\ \bibinfo
  {pages} {534} (\bibinfo {year} {1998})}\BibitemShut {NoStop}%
\bibitem [{\citenamefont {Livadiotis}\ and\ \citenamefont
  {McComas}(2009)}]{LivadiotisM09}%
  \BibitemOpen
  \bibfield  {author} {\bibinfo {author} {\bibfnamefont {G.}~\bibnamefont
  {Livadiotis}}\ and\ \bibinfo {author} {\bibfnamefont {D.~J.}\ \bibnamefont
  {McComas}},\ }\href {\doibase 10.1002/10.1019/2009JA014352} {\bibfield
  {journal} {\bibinfo  {journal} {J. Geophys. Res.}\ }\textbf {\bibinfo
  {volume} {114}},\ \bibinfo {pages} {A11105} (\bibinfo {year}
  {2009})}\BibitemShut {NoStop}%
\bibitem [{\citenamefont {Ziebell}\ \emph
  {et~al.}(2014{\natexlab{a}})\citenamefont {Ziebell}, \citenamefont {Yoon},
  \citenamefont {Jr.}, \citenamefont {Gaelzer},\ and\ \citenamefont
  {Pavan}}]{pl:ZiebellYSGP14}%
  \BibitemOpen
  \bibfield  {author} {\bibinfo {author} {\bibfnamefont {L.~F.}\ \bibnamefont
  {Ziebell}}, \bibinfo {author} {\bibfnamefont {P.~H.}\ \bibnamefont {Yoon}},
  \bibinfo {author} {\bibfnamefont {F.~J. R.~S.}\ \bibnamefont {Jr.}}, \bibinfo
  {author} {\bibfnamefont {R.}~\bibnamefont {Gaelzer}}, \ and\ \bibinfo
  {author} {\bibfnamefont {J.}~\bibnamefont {Pavan}},\ }\href {\doibase
  10.1063/1.4861619} {\bibfield  {journal} {\bibinfo  {journal} {Phys.
  Plasmas}\ }\textbf {\bibinfo {volume} {21}},\ \bibinfo {pages} {010701}
  (\bibinfo {year} {2014}{\natexlab{a}})}\BibitemShut {NoStop}%
\bibitem [{\citenamefont {Ziebell}\ \emph
  {et~al.}(2014{\natexlab{b}})\citenamefont {Ziebell}, \citenamefont {Yoon},
  \citenamefont {Gaelzer},\ and\ \citenamefont {Pavan}}]{pl:ZiebellYGP14}%
  \BibitemOpen
  \bibfield  {author} {\bibinfo {author} {\bibfnamefont {L.~F.}\ \bibnamefont
  {Ziebell}}, \bibinfo {author} {\bibfnamefont {P.~H.}\ \bibnamefont {Yoon}},
  \bibinfo {author} {\bibfnamefont {R.}~\bibnamefont {Gaelzer}}, \ and\
  \bibinfo {author} {\bibfnamefont {J.}~\bibnamefont {Pavan}},\ }\href
  {\doibase 10.1063/1.4863453} {\bibfield  {journal} {\bibinfo  {journal}
  {Phys. Plasmas}\ }\textbf {\bibinfo {volume} {21}},\ \bibinfo {pages}
  {012306} (\bibinfo {year} {2014}{\natexlab{b}})}\BibitemShut {NoStop}%
\bibitem [{\citenamefont {Newbury}\ \emph {et~al.}(1998)\citenamefont
  {Newbury}, \citenamefont {Russell}, \citenamefont {Phillips},\ and\
  \citenamefont {Gary}}]{Newbury1998}%
  \BibitemOpen
  \bibfield  {author} {\bibinfo {author} {\bibfnamefont {J.~A.}\ \bibnamefont
  {Newbury}}, \bibinfo {author} {\bibfnamefont {C.~T.}\ \bibnamefont
  {Russell}}, \bibinfo {author} {\bibfnamefont {J.~L.}\ \bibnamefont
  {Phillips}}, \ and\ \bibinfo {author} {\bibfnamefont {S.~P.}\ \bibnamefont
  {Gary}},\ }\href {\doibase 10.1029/98JA00067} {\bibfield  {journal} {\bibinfo
   {journal} {Journal of Geophysical Research: Space Physics}\ }\textbf
  {\bibinfo {volume} {103}},\ \bibinfo {pages} {9553} (\bibinfo {year}
  {1998})}\BibitemShut {NoStop}%
\bibitem [{\citenamefont {Ziebell}\ \emph
  {et~al.}(2014{\natexlab{c}})\citenamefont {Ziebell}, \citenamefont {Yoon},
  \citenamefont {Gaelzer},\ and\ \citenamefont {Pavan}}]{pl:ZiebellYGP14b}%
  \BibitemOpen
  \bibfield  {author} {\bibinfo {author} {\bibfnamefont {L.~F.}\ \bibnamefont
  {Ziebell}}, \bibinfo {author} {\bibfnamefont {P.~H.}\ \bibnamefont {Yoon}},
  \bibinfo {author} {\bibfnamefont {R.}~\bibnamefont {Gaelzer}}, \ and\
  \bibinfo {author} {\bibfnamefont {J.}~\bibnamefont {Pavan}},\ }\href
  {\doibase 10.1088/2041-8205/795/2/L32} {\bibfield  {journal} {\bibinfo
  {journal} {Astrophys. J. Lett.}\ }\textbf {\bibinfo {volume} {795}},\
  \bibinfo {pages} {L32} (\bibinfo {year} {2014}{\natexlab{c}})}\BibitemShut
  {NoStop}%
\bibitem [{\citenamefont {Ziebell}\ \emph {et~al.}(2015)\citenamefont
  {Ziebell}, \citenamefont {Yoon}, \citenamefont {Petruzzellis}, \citenamefont
  {Gaelzer},\ and\ \citenamefont {Pavan}}]{pl:ZiebellYPGP15}%
  \BibitemOpen
  \bibfield  {author} {\bibinfo {author} {\bibfnamefont {L.~F.}\ \bibnamefont
  {Ziebell}}, \bibinfo {author} {\bibfnamefont {P.~H.}\ \bibnamefont {Yoon}},
  \bibinfo {author} {\bibfnamefont {L.~T.}\ \bibnamefont {Petruzzellis}},
  \bibinfo {author} {\bibfnamefont {R.}~\bibnamefont {Gaelzer}}, \ and\
  \bibinfo {author} {\bibfnamefont {J.}~\bibnamefont {Pavan}},\ }\href
  {\doibase 10.1088/0004-637X/806/2/237} {\bibfield  {journal} {\bibinfo
  {journal} {Astrophys. J.}\ }\textbf {\bibinfo {volume} {806}},\ \bibinfo
  {eid} {237} (\bibinfo {year} {2015})}\BibitemShut {NoStop}%
\end{thebibliography}%

\end{document}